\documentclass[preprint,preprintnumbers,aps,superscriptaddress,nofootinbib,tightenlines,floatfix]{revtex4}

\usepackage{amsmath,amssymb}
\usepackage{graphicx}
\usepackage{bm}
\usepackage{comment}

\newcommand{\beq}{\begin{equation}}
\newcommand{\eeq}{\end{equation}}
\newcommand{\bea}{\begin{eqnarray}}
\newcommand{\eea}{\end{eqnarray}}

\newcommand{\gsim}{\raisebox{-0.7ex}{$\stackrel{\textstyle >}{\sim}$ }}

\def\Nprops{0.435\times 10^6} 
\def\NpropsOLD{0.28\times 10^6} 
\def\Ncfgs{1195}
\def\NcfgsOLD{1194}
\def\PropsperCFG{364}

\def\mpi{m_\pi}

\def\OMIT#1{{}}

\newcommand{\lsim}{\raisebox{-0.7ex}{$\stackrel{\textstyle <}{\sim}$ }}

\newcount\hour \newcount\hourminute \newcount\minute 
\hour=\time \divide \hour by 60
\hourminute=\hour \multiply \hourminute by 60
\minute=\time \advance \minute by -\hourminute
\newcommand{\mydate}{\ \today \ - \number\hour :\number\minute}

\begin{document}

\begin{figure}[!t]

  \vskip -1.5cm
  \leftline{\includegraphics[width=0.25\textwidth]{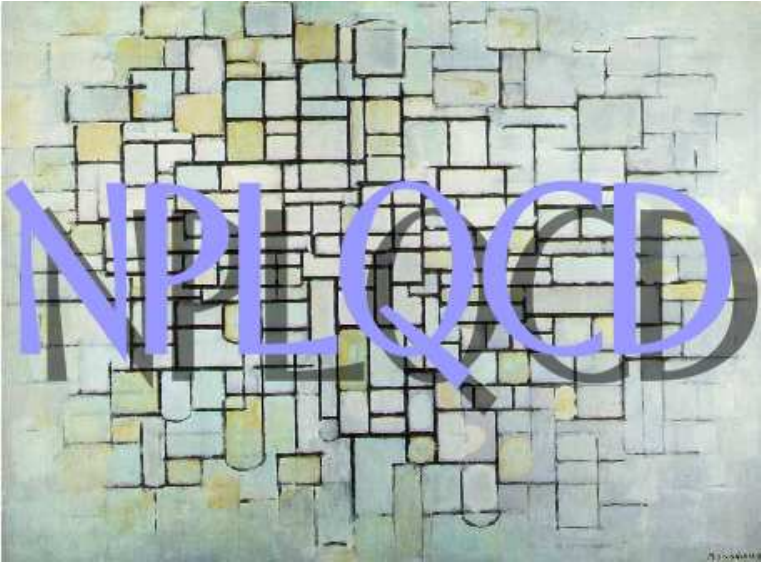}}
\end{figure}

\preprint{\vbox{ \hbox{UNH-09-06} \hbox{JLAB-THY-09-1116}
    \hbox{NT@UW-09-26}
    \hbox{IUHET-539}\hbox{ATHENA-PUB-09-019} }}

\title{High Statistics Analysis using Anisotropic Clover Lattices:
  (III) Baryon-Baryon Interactions}

\author{Silas R.~Beane} \affiliation{Department of Physics, University
  of New Hampshire, Durham, NH 03824-3568.}  \author{William Detmold}
\affiliation{Department of Physics, College of William and Mary,
  Williamsburg, VA 23187-8795.}  \affiliation{Jefferson Laboratory,
  12000 Jefferson Avenue, Newport News, VA 23606.}  \author{Huey-Wen
  Lin} \affiliation{Department of Physics, University of Washington,
  Seattle, WA 98195-1560.}  \author{Thomas C.~Luu} \affiliation{N
  Division, Lawrence Livermore National Laboratory, Livermore, CA
  94551.}  \author{Kostas Orginos} \affiliation{Department of Physics,
  College of William and Mary, Williamsburg, VA 23187-8795.}
\affiliation{Jefferson Laboratory, 12000 Jefferson Avenue, Newport
  News, VA 23606.} 
 \author{Martin J.~Savage}
\affiliation{Department of Physics, University of Washington, Seattle,
  WA 98195-1560.} 
 \author{Aaron Torok} \affiliation{Department of
  Physics, Indiana University, Bloomington, IN 47405.}
\author{Andr\'e Walker-Loud} \affiliation{Department of Physics,
  College of William and Mary, Williamsburg, VA 23187-8795.}
\collaboration{ NPLQCD Collaboration } \noaffiliation \vphantom{}

\date{\mydate}

\begin{abstract}
  \noindent
  Low-energy baryon-baryon interactions are calculated in a
  high-statistics lattice QCD study on a single ensemble of
  anisotropic clover gauge-field configurations at a pion mass of
  $m_\pi\sim 390$ MeV, a spatial volume of $L^3\sim (2.5\ {\rm
    fm})^3$, and a spatial lattice spacing of $b\sim 0.123~{\rm fm}$.
  L\"uscher's method is used to extract nucleon-nucleon,
  hyperon-nucleon and hyperon-hyperon scattering phase shifts at one
  momentum from the one- and two-baryon ground-state energies in the
  lattice volume.  The isospin-3/2 $N\Sigma$ interactions are found to be highly
  spin-dependent, and the interaction in the $^3S_1$ channel is found
  to be strong.  In contrast, the $N\Lambda$ interactions are found to
  be spin-independent, within the uncertainties of the calculation,
  consistent with the absence of one-pion-exchange.  The only channel
  for which a negative energy-shift is found is $\Lambda\Lambda$,
  indicating that the $\Lambda\Lambda$ interaction is attractive, as
  anticipated from model-dependent discussions regarding the
  H-dibaryon.  The NN scattering lengths are found to be small,
  clearly indicating the absence of any fine-tuning in the NN-sector
  at this pion mass. This is consistent with our previous Lattice QCD
  calculation of NN interactions.  The behavior of the signal-to-noise
  ratio in the baryon-baryon correlation functions, and in the ratio
  of correlation functions that yields the ground-state energy
  splitting is explored.  In particular, focus is placed on the window
  of time slices for which the signal-to-noise ratio does not degrade
  exponentially, as this provides the opportunity to extract
  quantitative information about multi-baryon systems.
\end{abstract}
\maketitle

%
%
\section{Introduction \label{sec:Intro} }

\noindent The strong interactions among baryons are key to every
aspect of our existence.  The two- and higher-body interactions among
protons and neutrons conspire to produce the spectrum of nuclei and
the complicated chains of nuclear reactions that allow for the
production of the elements forming the periodic table at the earliest
times of our universe, in the stellar environments that follow, and in
reactors and our laboratories.  Decades of experimental effort have
provided a very-precise set of measurements of the nucleon-nucleon
scattering cross sections over a wide range of
energies~\cite{NNonline}, and these cross sections have in turn given
rise to the modern nuclear forces.  These experimentally determined
two-body forces, encoded by potentials such as ${\rm
  AV}_{18}$~\cite{Wiringa:1994wb} and the chiral
potentials~\cite{Epelbaum:2008ga}, when supplemented with three-body
interactions, now provide the cornerstone of our theoretical
description of nuclei~\cite{Pieper:2007ax} and their interactions.
The two-nucleon forces are observed to be significantly more important
than the three-nucleon forces, which, in turn, are significantly more
important than the four-nucleon forces.  Present-day calculations are
sufficiently precise that the inclusion of three-nucleon forces is
required in order to post-dict the structure of light
nuclei~\cite{Pieper:2004qw,Navratil:2007we}.  Given the relatively
small contributions of the three-nucleon and higher-body interactions
to light-nuclei, where reliable calculations are presently possible,
there is considerable uncertainty with regard to their form.  This
leads to enhanced uncertainties in the calculation of systems for
which there is little or no experimental guidance, such as moderate to
high-density neutron-rich environments, and more generally, nuclear
environments at densities exceeding that of nuclear matter.

In dense nuclear systems it is not only the multi-nucleon forces that
are difficult to quantify as non-nucleonic objects may play an
important role.  In a core-collapse supernova, it is the nuclear
equation of state (NEOS) that ultimately dictates whether the system
collapses into a neutron star or forms a black hole.  This, in turn,
is determined by the composition and structure of the hadronic matter
in the core of the supernova, which is at (baryon number) densities
that are a few times that of nuclear matter.  At such densities, the
strange quark may play a pivotal role through the formation of a
charged kaon condensate made possible by the strength of attractive
kaon-nucleon interactions. It may also become energetically favorable
for the matter to contain strange baryons, such as the $\Sigma^-$ or
$\Lambda$, because of their interactions with nucleons.  In either
case, it is the two-body interactions between the strange hadrons and
the nucleons, and also between themselves, that largely determine the
composition of the hadronic matter at core-collapse densities, and
ultimately the fate of the collapsing supernova~\cite{Page:2006ud}.
Unfortunately, experimental determinations of the interactions of
strange hadrons are very challenging due to their weak decays, and
existing cross-section measurements are not precise enough to provide
meaningful constraints on the NEOS at core-collapse densities.  So,
while it is the interactions between three-nucleons that currently
limit the precision with which properties of material composed of
neutrons and protons can be calculated, it is the two-body
interactions involving strange hadrons that currently provide the most
serious limitation to reliable calculations at densities that exceed
that of nuclear matter.

One of the major objectives of lattice QCD (LQCD) is to calculate the
properties and interactions of nucleons and, more generally, systems
comprised of multiple hadrons.  Precise exploration of the simplest
multi-hadron systems has recently become possible with significant
advances in computational resources, as well as through algorithmic
and theoretical developments (for a recent review, see
Ref~\cite{Beane:2008dv}).  The $I=2$ two-pion system, $\pi^+\pi^+$, is
the simplest of such multi-hadron systems to calculate in LQCD, and
current computational resources have allowed for a $\sim 1\% $ level
determination of the $\pi^+\pi^+$ scattering
length~\cite{Beane:2005rj,Beane:2007xs,Feng:2009ij}.  Further, systems
comprised of up to twelve $\pi^+$'s~\cite{Beane:2007es,Detmold:2008fn}
and systems comprised of up to twelve $K^+$'s~\cite{Detmold:2008yn}
have been explored, allowing a determination of the three-$\pi^+$ and
three-$K^+$ interactions as well as aspects of pion and kaon
condensates.

In general, a determination of the two-particle scattering amplitude,
or multi-body interactions, with LQCD requires calculating the energy
eigenvalues of the appropriate two-hadron system in the finite
volume~\cite{Hamber:1983vu,Luscher:1986pf,Luscher:1990ux}.  The energy
differences between the multi-particle energy levels in the finite
volume and the sum of the particle masses determines the scattering
amplitude at the corresponding energy.  Processes of interest to
low-energy nuclear physics occur in the MeV energy regime, while the
masses of the baryons and nuclei are in the GeV regime. As a result,
very precise (high-statistics) measurements must be performed in order
to reliably extract useful constraints on two- and higher-body
interactions from a lattice calculation.

In contrast to mesonic systems, lattice QCD calculations in the baryon
sector are complicated by the statistical behavior of the correlation
functions~\cite{Lepage:1989hd}.  We performed the first unquenched
LQCD calculation of nucleon-nucleon (NN)~\cite{Beane:2006mx} and
hyperon-nucleon (YN)~\cite{Beane:2006gf} scattering lengths, and found
that the NN scattering lengths appear to be of natural size at larger
pion masses, and are set by the range of the interaction.
Consequently, an increase in the magnitude of the scattering length as
the pion mass is reduced down to its physical value is anticipated.
Further, we have recently performed the first lattice QCD calculation
of scattering in various meson-baryon channels~\cite{Torok:2009dg}
using the mixed-action scheme of domain-wall valence quarks on the
MILC staggered sea.

Recently, we have performed a high-statistics calculation of a number
of single-hadron correlation functions~\cite{Beane:2009ky} and also
the first calculations of three-baryon systems~\cite{Beane:2009gs} on
an ensemble of the anisotropic gauge-field configurations generated by
the Hadron Spectrum
Collaboration~\cite{Lin:2008pr,Edwards:2008ja}. This ensemble has a
pion mass of $m_\pi\sim 390~{\rm MeV}$, a spatial lattice spacing of
$b_s= 0.1227\pm 0.0008~{\rm fm}$, an anisotropy $\xi=b_s/b_t=3.500\pm
0.032$ and a lattice volume of $L^3\times T=20^3\times 128$.  The goal
of those studies was to ``jump'' an order of magnitude in the number
of measurements performed to estimate correlation functions, and to
explore the ``new territory'' that subsequently emerged.  The baryon
masses were extracted with fully quantified uncertainties at the
$\lsim 0.2\%$-level from the $\NpropsOLD$ measurements performed on
$\NcfgsOLD$ gauge-field configurations. With a somewhat smaller
statistical sample, the binding energies of the $pnn$ (triton $\equiv$
$^3$He by the isospin symmetry of the LQCD calculation) and
``$\Xi^0\Xi^0n$'' three-baryon systems were investigated and in the
latter case, determined with an uncertainty of $\sim 3$
MeV/baryon. Subsequently, a study of the $^3$He ($ppn$) and $^4$He
($ppnn$) systems in quenched QCD at a large quark mass appeared
\cite{Yamazaki:2009ua}. A number of important and surprising
observations were made in those
works~\cite{Beane:2009ky,Beane:2009gs,Yamazaki:2009ua} that have
modified how we foresee moving toward calculating the properties of
light nuclei.

One of the most important aspects of our previous work was the
detailed study of the signal-to-noise ratio in the single-baryon
correlation functions.  The signal-to-noise ratio was found to be
approximately independent of time for a significant number of
time-slices~\cite{Beane:2009ky,Beane:2009gs} prior to evolving toward
the expected exponential degradation~\cite{Lepage:1989hd}.  This
window of time-slices is understood in terms of the suppression of
contributions from purely mesonic states in the correlation function
that determines the variance of the single-baryon correlation
function.  Given that the signal-to-noise ratio for a system
containing more than one baryon is expected to scale (approximately)
as the product of the signal-to-noise ratio's of the individual
baryons (neglecting their interactions), this window of time slices
suggests that calculations of the energy levels of systems containing
a number of baryons in this lattice volume with these interpolating
operators is possible. In this work we present the baryon-baryon
scattering phase shifts that have been measured in our high-statistics
anisotropic clover-quark calculation.

Our analysis and results are presented in the following manner. In
Section \ref{sec:latt-qcd-calc}, we introduce the details of the
formalism used in extracting phase shifts from our LQCD
calculations. Section \ref{sec:baryons} briefly reviews our single
baryon results before we present our extractions of two-baryon
interactions in Section \ref{sec:bary-bary-inter}. Section
\ref{sec:stat-behav} details our analysis of statistical scaling and
noise in the various correlation functions while Section
\ref{sec:discussion} is a concluding discussion.

\section{Lattice QCD calculations}
\label{sec:latt-qcd-calc}

\subsection{L\"uscher method for extracting scattering parameters}
\label{sec:luscher-method}

\noindent In this work, the finite volume scaling method (L\"uscher's
method)~\cite{Hamber:1983vu,Luscher:1986pf,Luscher:1990ux} is employed
to extract the two-particle scattering amplitudes below inelastic
thresholds at a given energy.  In the situation where only a single
scattering channel is kinematically allowed, the deviation of the
energy eigenvalues of the two-hadron system in the lattice volume from
the sum of the single-hadron masses is related to the scattering phase
shift, $\delta$.  For energy eigenvalues above kinematic thresholds
where multiple channels contribute, a coupled-channels analysis is
required as a single phase shift does not parameterize the S-matrix.
The energy shift for two particles $A$ and $B$, $\Delta E = E_{AB} -
E_A - E_B$, can be determined from the correlation functions for
systems containing one and two hadrons.  For baryon-baryon systems,
correlation functions of the form
\begin{eqnarray}
  \label{eq:4}
  C_{{\cal B};\Gamma}({\bf p},t) &=& \sum_{\bf x} \ e^{i{\bf p}\cdot{\bf x}} \ 
  \Gamma_{\alpha}^\beta\ 
  \langle {\cal B}_\alpha({\bf x},t)\  \overline{\cal
    B}_\beta({\bf x}_0,0)\rangle
  \\
  C_{{\cal B}_1,{\cal B}_2;\Gamma}({\bf p}_1,{\bf p}_2,t) 
  &=& \sum_{{\bf x}_1,{\bf
      x}_2} e^{i{\bf p}_1\cdot{\bf x}_1} 
  e^{i{\bf p}_2\cdot{\bf x}_2} 
  \Gamma_{\beta_1\beta_2}^{\alpha_1\alpha_2}
  \langle {\cal B}_{1,\alpha_1}({\bf x}_1,t){\cal B}_{2,\alpha_2}({\bf x}_2,t) 
  \overline{\cal
    B}_{1,\beta_1}({\bf x}_0,0) \overline{\cal
    B}_{2,\beta_2}({\bf x}_0,0) \rangle \,,\nonumber 
\end{eqnarray}
are used, where ${\cal B}$ denotes a baryon interpolating operator,
$\alpha_{(i)}$ and $\beta_{(i)}$ are Dirac indices, and the $\Gamma$
are spin tensors that typically project onto particular parity and/or
angular momentum states.  The baryon octet interpolating operators are
of the form
\begin{eqnarray}
  \label{eq:7}
  p_\alpha({\bf x},t) &=& \epsilon^{ijk} u_\alpha^i({\bf
    x},t)\left(u^{j {\sf T}}({\bf x},t)C\gamma_5 d^k({\bf
      x},t)\right)\,,
  \nonumber \\
  \Lambda_\alpha({\bf x},t) &=& \epsilon^{ijk} s_\alpha^i({\bf
    x},t)\left(u^{j {\sf T}}({\bf x},t)C\gamma_5 d^k({\bf x},t)\right)\,,
  \nonumber \\
  \Sigma^+_\alpha({\bf x},t) &=& \epsilon^{ijk} u_\alpha^i({\bf
    x},t)\left(u^{j {\sf T}}({\bf x},t)C\gamma_5 s^k({\bf x},t)\right)\,,
  \nonumber \\
  \Xi^0_\alpha({\bf x},t) &=& \epsilon^{ijk} s_\alpha^i({\bf
    x},t)\left(u^{j {\sf T}}({\bf x},t)C\gamma_5 s^k({\bf x},t)\right)\,,
\end{eqnarray}
where $C$ is the charge-conjugation matrix and $ijk$ are color
indices. Other hadrons in the lowest-lying octet can be obtained from
the appropriate combinations of quark flavors.  The parentheses in the
interpolating operators indicate contraction of spin indices into a
spin-0 ``diquark''.  It is worth pointing out that the overlap of the
composite operator ${\cal B}_{1,\alpha_1}({\bf x}_1,t){\cal
  B}_{2,\alpha_2}({\bf x}_2,t) $ onto two-baryon states is not simply
the product of individual baryon ``$Z$-factors'', but depends
explicitly upon ${\bf x}_1 - {\bf x}_2$, with a correlation length set
by the pion mass.  This fact precludes a determination of
interpolator-independent baryon-baryon potentials from lattice QCD
calculations \cite{Detmold:2007wk}.

Away from the time slice on which the source is placed (in this case
$t=0$) these correlation functions behave as
\begin{eqnarray}
  \label{eq:5}
  C_{{\cal H}_A}({\bf p},t)
  & = & 
  \sum_n \ Z_{n;A}^{(i)}({\bf p}) \ Z_{n;A}^{(f)}({\bf p}) 
  \ e^{- E_n^{(A)}({\bf p})\  t} \,, \\
  C_{{\cal H}_A{\cal H}_{B}}({\bf p},-{\bf p},t)
  & = &  \sum_n\  Z_{n;AB}^{(i)}({\bf p}) \  Z_{n;AB}^{(f)}({\bf p}) 
  \ e^{- E_{n}^{(AB)}({\bf 0})\  t} \,,
\end{eqnarray}
with $E_0^{(A)}({\bf 0})=m_A$ and the $E_n^{(AB)}({\bf 0})$ are the
energy-eigenvalues of the two particle system (we only present
calculations of two-baryon systems for which the center-of-mass is at
rest) in the lattice volume.  At large times, the ratio
\begin{eqnarray}
  \label{eq:6}
  \frac{C_{{\cal H}_A{\cal H}_{B}}({\bf p},-{\bf p},t)}{C_{{\cal H}_A}({\bf
      0},t)C_{{\cal H}_{B}}({\bf 0},t)}
  &\stackrel{t\to\infty}{\longrightarrow}&  
  \widetilde{Z}_{0,AB}^{(i)}({\bf p})\widetilde{Z}_{0,AB}^{(f)}({\bf p})
  \ e^{- \Delta E_{0}^{(AB)}({\bf 0})\    t}
  \,
\end{eqnarray}
decays as a single exponential in time with the energy shift, $\Delta
E_{0}^{(AB)}({\bf 0})$.  In what follows, only the case ${\bf p}={\bf
  0}$ is considered.  The energy shifts
\begin{eqnarray}
  \label{eq:3}
  \Delta E_n^{(AB)} \equiv \Delta E_n^{(AB)}({\bf 0}) 
  &\equiv& E_n^{(AB)}({\bf 0}) - m_A - m_B = \sqrt{q_n^2 + m_A^2} +
  \sqrt{q_n^2 + m_B^2} -m_A -m_B
  \nonumber \\
  &=&\frac{q_n^2}{2\mu_{AB}}+ \ldots\,,
\end{eqnarray}
(where $\mu_{AB}=m_A m_B/(m_A+m_B)$ is the reduced mass of the
two-particle system) determine squared momenta, $q_n^2$ (which can be
either positive or negative). Below inelastic thresholds, these are
related to the real part of the inverse scattering amplitude
via~\footnote{Calculations performed on anisotropic lattices, such as
  those used in this work, require a modified energy-momentum
  relation, and as a result eq.~(\ref{eq:3}) becomes
  \begin{eqnarray}
    \label{eq:3xi}
    \Delta E_n^{(AB)} &\equiv& E_n^{(AB)} - m_A - m_B = 
    \sqrt{q_n^2/\xi_A^2 + m_A^2} +
    \sqrt{q_n^2/\xi_B^2 + m_B^2} -m_A -m_B
    \ \ ,
  \end{eqnarray}
  where $\xi_{A,B}$ are the anisotropy factors for particle $A$ and
  particle $B$, respectively, determined from the appropriate
  energy-momentum dispersion relation.  The masses and energy
  splitting are given in terms of temporal lattice units and $q_n$ is
  given in spatial lattice units.  In the present work we find that
  the various $\xi_A$ agree within uncertainties and have used
  $\xi_A=\xi_B=\xi$ for all scattering processes.  }
\begin{equation}
  \label{eq:1}
  q_n\, \cot \delta(q_n) = 
  \frac{1}{\pi\ L} S\left(q_n^2 \left(\frac{L}{2\pi}\right)^2\right)\,,
\end{equation}
where
\begin{equation}
  \label{eq:2}
  S(x)=\lim_{\Lambda\to\infty} \sum_{\bf j}^{|{\bf
      j}|<\Lambda}\frac{1}{|{\bf j}|^2 - x}  -4\pi\ \Lambda\,,
\end{equation}
thereby implicitly determining the value of the phase shift at the
energy $\Delta E_n^{(AB)}$ (or center of mass momentum $q_n$),
$\delta(q_n)$.  Thus, the function $p\cot\delta(p)$, that determines
the low-energy elastic-scattering cross-section, ${\cal
  A}(p)\propto(p\cot\delta(p)-i\,p)^{-1}$, is determined at the energy
$\Delta E_n^{(AB)}$.

For a scattering process for which the exchange of a single pion
(One-Pion-Exchange (OPE)) is allowed by spin and isospin
considerations, the function $p\cot\delta(p)$ is a analytic function
of $|{\bf p}|^2$ for $|{\bf p}|\leq m_\pi/2$ (determined by the
t-channel cut in the scattering amplitude).  In this kinematic regime,
$p\cot\delta(p)$ has a series-expansion (the effective range
expansion) of the form
\begin{eqnarray}
  p\cot\delta(p) & = & 
  -{1\over a}\ +\ {1\over 2}\  r_0\ |{\bf p}|^2\ +\ ...
  \ \ \ ,
  \label{eq:ere}
\end{eqnarray} 
where $a$ is the scattering length (with the nuclear physics sign
convention) and $r_0$ is the effective range.  While the magnitude of
the effective range (and higher terms) is set by the pion mass, the
scattering length is unconstrained.  For scattering processes where
OPE is not allowed, the lower limit of the cut in the t-channel and
the location of inelastic threshold set the radius of convergence of
the effective-range expansion of $p\cot\delta(p)$.

\subsection{Expectations at finite temporal extent}
\label{sec:expect-at-finite-1}

\noindent
For baryon-baryon correlation functions computed from quark
propagators that are anti-periodic in the time direction, and using
the sources of eq.~(\ref{eq:7}), the correlation functions contain
contributions from hadronic states propagating forwards and backwards
in time. They are expected to have the form
\begin{eqnarray}
  \label{eq:9}
  C_{{\cal BB};\Gamma_+}({\bf 0},{\bf 0},t) & = & 
  Z_1\ e^{-E_{\cal BB}t} 
  \ +\ 
  Z_2\ e^{-E_{ {\cal B}^\prime {\cal B}^\prime } (T-t)}
  \ +\ 
  Z_3\ e^{-E_{\cal B} t}e^{-E_{{\cal B}^\prime}(T-t)} 
  \nonumber\\
  && + \ 
  Z_4\ e^{-E_{\cal BB}(T-t)}
  +Z_5\ e^{-E_{{\cal B}^\prime {\cal B}^\prime}t}
  \ +\ 
  \ldots
  \,,
\end{eqnarray}
where $E_{\cal BB}$ is the energy of the two-baryon state, $E_{\cal
  B}$ is the ground-state baryon energy of positive parity, $E_{{\cal
    B}^\prime}$ is the energy of the ground-state negative-parity
baryon or meson-baryon scattering state and the ellipsis denotes
contributions from higher excited states of the same quantum numbers.
The exponent of the first term in eq.~(\ref{eq:9}) is the primary
object of our study and the other contributions are viewed as
``pollutants''.

Since the interpolating operators for the baryons that give rise to
the baryon-baryon correlation functions are individually projected
with the positive-energy projectors, $\Gamma_\pm=(1\pm\gamma_0)/2$,
the $Z$-factors of the forward and backward propagating states of the
same energy in eq.~(\ref{eq:9}) are not related, for example it is
possible that $Z_4\ne Z_1$.  Indeed, since the sink separately
projects the parity of each baryon (along with its momentum), the
amplitude of the two backwards propagating baryons is suppressed by
the lattice volume relative to the two forwards propagating
baryons. This is a significant suppression, and we expect to see more
energetic states, such as ${\cal B}^\prime {\cal B}^\prime$,
dominating the correlation function at large times near the boundary.

\subsection{Computational overview}
\label{sec:comp-deta}
\noindent
In the present study, we have employed a single ensemble of
$20^3\times 128$ gauge-field configurations generated with a $n_f=2+1$
flavor, anisotropic-clover quark action, with a spatial lattice
spacing of $b= 0.1227(8)~{\rm fm}$ and a pion mass $m_\pi\sim 390~{\rm
  MeV}$, that have been produced by the Hadron Spectrum
Collaboration~\cite{Lin:2008pr,Edwards:2008ja}.  The technical details
of the propagators computed on this ensemble are presented in
Ref.~\cite{Beane:2009ky} and we do not repeat them here. In the
current calculation, the number of measurements has been increased to
an average of $\PropsperCFG$ randomly-distributed measurements on each
of $\Ncfgs$ configurations (a total of $\sim\Nprops$
measurements). For correlators from each source point, two types of
sink interpolating operators are used \cite{Beane:2009ky} and the
resulting correlation functions are referred to as ``smeared-point''
(SP) and ``smeared-smeared'' (SS).  The measurements on each
configuration are averaged, and then these averaged measurements are
typically blocked (averaged) in sets of ten neighboring configurations
(100 trajectories) to account for residual correlations (see
Ref.~\cite{Beane:2009ky} for a detailed study of correlations between
different sources and configurations).

The methods used to determine quantities of interest from the measured
correlation functions are discussed in detail in
Refs.~\cite{Beane:2009ky,Beane:2009gs}, and we present no more than an
overview here.  One method of extraction is a direct analysis of
``effective plots''.  Linear combinations of the SP and SS correlation
functions associated with each baryon and baryon-baryon state are
formed to eliminate the contribution from excited states, as discussed
in Ref.~\cite{Beane:2009ky}.  Effective mass plots are formed from
these correlation functions and ratios of correlation functions to
extract the energy and energy splitting, respectively\footnote{ Linear
  combination of SP and SS correlation functions are formed that
  extend the plateau of the ground state found from a matrix-Prony
  analysis \cite{Beane:2009gs} to earlier times.  The effect of
  choosing a slightly different linear combination is taken into
  account by the systematic error. While we find that the best
  candidate for the ground state level is not ambiguous in any of the
  two-baryon correlation functions, there is always the possibility
  that the ``true'' ground state becomes dominant in the region where
  it is likely that there are significant contributions from
  backward-propagating states which contaminate the
  signal. Eliminating this possibility requires better statistical
  precision and/or measurements on lattices with a longer time
  extent.}. Time intervals over which the effective mass appears
constant are identified, and the energy is extracted from a correlated
$\chi^2$-squared minimization with the covariance matrix determined
with the Jackknife or Bootstrap procedure.  A fitting systematic
uncertainty is assigned from the range of extracted energies
determined by varying the location and size of the fitting interval
over a reasonable range.  A given energy-splitting and its uncertainty
is converted into a value of $q_n^2$ and its uncertainty using
eq.~(\ref{eq:3}), or a direct fit to an effective-$q_n^2$ plot is
performed, which is then translated into $q_n\cot\delta(q_n)$ and its
associated uncertainty using eq.~(\ref{eq:1}).  There are a number of
ways to perform this last stage of the analysis. One way is to use the
Jackknife or Bootstrap procedure to determine the uncertainty in
$q_n\cot\delta(q_n)$ directly.  However, this is complicated by the
fact that $S(x)$ in eq.~(\ref{eq:1}) is a singular function.  An
alternate method is to determine the value of $q_n^2$ and its
associated uncertainty from the two-baryon energy splitting from
eq.~(\ref{eq:3}), and then to propagate the central value and
uncertainties through eq.~(\ref{eq:1}) to determine
$q_n\cot\delta(q_n)$.  We present results using the latter method.

Results from this methodology are consistent with those from
multi-exponential fits to the correlation functions and with various
matrix-Prony \cite{Beane:2009gs} based analyses. In the following, we
present results from a single analysis, ensuring that the systematic
uncertainties are sufficient to maintain agreement with analyses using
these other methods.

\section{Single Baryons}
\label{sec:baryons}

\begin{figure}[!th]
  \centering
  \includegraphics[width=1.0\columnwidth]{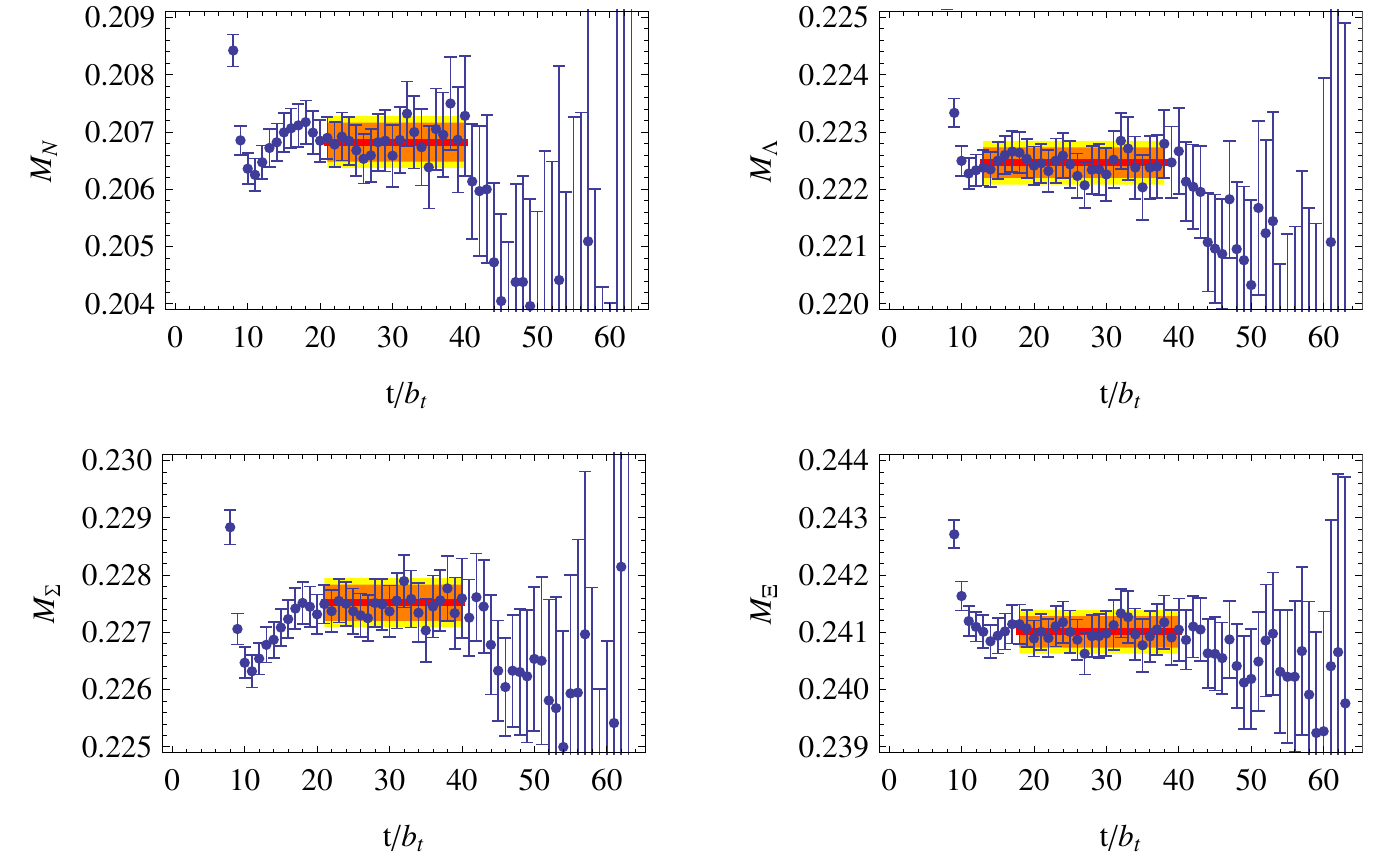}
  \caption{The single-baryon GEMP's (with $t_J=3$) resulting from the
    linear combination of SS and SP correlation functions that
    eliminates contributions from excited states.  The fit values of
    the masses along with the statistical uncertainty, and the
    systematic and statistical uncertainties combined in quadrature,
    are shown. }
  \label{fig:B-BOT}
\end{figure}

\noindent The interaction between baryons is extracted from the
difference between the energy-levels of the two baryon system in the
lattice volume and the individual baryon masses.  The masses of the
baryons were extracted in a previous work~\cite{Beane:2009ky}, but it
is useful to show the masses here, particularly due to the substantial
increase in the number of measurements that have been performed.  The
baryon masses are extracted from correlated $\chi^2$ minimizing fits
to the generalized effective mass plots (GEMP's) obtained from each
correlation function, $C_i(t)$, defined to be
\begin{eqnarray}
  M_{\rm eff ; t_J}(t) & = & 
  {1\over t_J}\ \log\left({C(t)\over C(t+t_J)}\right)
  \ \ \ .
  \label{eq:GEMP}
\end{eqnarray}
The GEMP's from the single-baryon correlation functions are presented
in fig.~\ref{fig:B-BOT}.
The extracted baryon masses that are fit to the plateau regions of the
GEMP's are shown along with the statistical uncertainty, and the
systematic and statistical uncertainties combined in quadrature.  The
results of the fitting, along with the fitting intervals are given in
Table~\ref{tab:baryonmasses}.
\begin{table}[!ht]
  \caption{Extracted single hadron masses.  
    A lattice spacing of $b_s=0.1227\pm 0.0008~{\rm fm}$
    and an anisotropy factor of $\xi_s= 3.500\pm 0.032$ is used to convert from 
    temporal lattice units (t.l.u.) to MeV.  
    The first two uncertainties are the statistical and systematic uncertainty of
    the extraction in temporal lattice units, while the third uncertainty quoted
    for quantities in physical units is the combined lattice spacing and anisotropy
    uncertainty.
  }
  \label{tab:baryonmasses}
  \begin{ruledtabular}
    \begin{tabular}{ccccc}
      Hadron  &  $M$ (t.l.u.) &  $M$ (MeV) &  $\chi /{\rm dof}$ & fitting interval  \\
      \hline
      $\pi$ & $0.06936(12)(05)$ & $390.39(0.67)(0.28)(4.38)$   & 0.73 & $21\rightarrow 41$\\
      K & $0.097016(99)(33)$ & $546.06(0.56)(0.19)(6.13)$   & 1.01 & $29\rightarrow 49$\\
      \hline
      N & $0.20682(34)(30)$ & $1164.1(1.9)(1.7)(13.1)$ & $1.16$ & $21\rightarrow 40$ \\
      $\Lambda$ & $0.22246(27)(27)$ & $1252.1(1.5)(1.5)(14.1)$ & $0.97$ & $13\rightarrow 38$ \\
      $\Sigma $ & $0.22752(32)(29)$ & $1280.6(1.7)(1.6)(14.3)$  & $1.46$ & $21\rightarrow 40$ \\
      $\Xi$ &  $0.24101(27)(27)$ & $1356.5(1.5)(1.5)(15.2)$ & $1.06$ & $16\rightarrow 40$\\
    \end{tabular}
  \end{ruledtabular}
\end{table}

The statistical and fitting systematic uncertainties are below 0.2\%
for each of the single baryon states (although the uncertainty in the
temporal lattice spacing leads to a larger uncertainty on the masses
in physical units).  The baryon signal-to-noise ratio is essentially
independent of time in the fitting windows shown in
Table~\ref{tab:baryonmasses} due to the nature of the sources that
generate the correlation function, as discussed in
Ref.~\cite{Beane:2009ky}.  It is only in these time intervals (for
this value of $t_J$) that reliable energy splittings between the
two-baryon and the single-baryon masses can be constructed from the
ratio of the correlation functions given in eq.~(\ref{eq:6}).

\section{Baryon-Baryon Interactions}
\label{sec:bary-bary-inter}

\noindent The energy eigenstates in the finite lattice volume are
classified by their global quantum numbers: baryon number, isospin,
third component of isospin, strangeness, total momentum, and behavior
under hyper-cubic transformations.  Six quark operators that are
simple products of three-quark baryon operators are used as sources
for the baryon-baryon correlation functions.  As a consequence, the
baryon content of the interpolating operator is used to define the
operator, e.g. $n\Lambda(^3S_1)$, but this operator will, in
principle, couple to all states in the volume with the quantum numbers
$B=2$, $I={1\over 2}$, $I_z=-{1\over 2}$, $s=-1$, and $^{2s+1}L_J =\;
^3S_1\ +\ ...$, where the ellipses denote states with higher total
angular momentum that also project onto the $A_1$ irreducible
representation of the cubic-group~\footnote{ The spatial dimensions of
  the gauge-field configurations that are used in this work are
  identical (i.e. isotropic), and as such the eigenstates of the QCD
  Hamiltonian can be classified with respect to their transformation
  properties under cubic transformations, $H(3)$, a subgroup of the
  group of continuous three-dimensional rotations, $O(3)$.  The
  two-baryon states that are calculated in this work all belong to the
  $A_1^+$ representation of $H(3)$, corresponding to combinations of
  states with angular momentum $L=0,4,6,\ldots$\ .}.  Both SS and SP
correlation functions have been calculated for the nine baryon-baryon
channels shown in Table~\ref{tab:channels}.
\begin{table}[!t]
  \caption{Baryon-baryon channels examined in this work.}
  \label{tab:channels}
  \begin{ruledtabular}
    \begin{tabular}{cccccc}
      & Channel   &  $I$ &  $|I_z|$ &  $s$  \\
      \hline 
      & $pp$ ($^1S_0$) & $1$ & $1$ & $0$  & \\
      & $np$ ($^3S_1$) & $0$ & $0$ & $0$  & \\
      & $n\Lambda$ ($^1S_0$) & ${1\over 2}$ & ${1\over 2}$ & $-1$  & \\
      & $n\Lambda$  ($^3S_1$) & ${1\over 2}$ & ${1\over 2}$ & $-1$  & \\
      & $n\Sigma^-$ ($^1S_0$) & ${3\over 2}$ & ${3\over 2}$ & $-1$  & \\
      & $n\Sigma^-$  ($^3S_1$) & ${3\over 2}$ & ${3\over 2}$ & $-1$ & \\
      & $\Sigma^-\Sigma^-$ ($^1S_0$) & $2$ & $2$ & $-2$  &\\
      & $\Lambda\Lambda$ ($^1S_0$) & $0$ & $0$ & $-2$  & \\
      & $\Xi^-\Xi^-$ ($^1S_0$) & $1$ & $1$ & $-4$  & \\
    \end{tabular}
  \end{ruledtabular}
\end{table}

If the calculations were performed on gauge-field configurations of
infinite extent in the time-direction, so that only forward
propagation could occur, some of the channels in
Table~\ref{tab:channels} could be analyzed by considering
contributions from a single scattering channel, e.g. $NN$,
$\Xi^-\Xi^-$, $\Sigma^-\Sigma^-$, $n\Sigma^-$, as we expect a single,
well-separated ground state for these quantum numbers. However, other
channels may require a multi-channel analysis, e.g. $n\Lambda$,
$\Lambda\Lambda$.  The $n\Lambda$ source will produce low-lying states
in the lattice volume that are predominately linear combinations of
the $n\Lambda$, $n\Sigma^0$ and $p\Sigma^-$ two-baryon states.  The
$\Lambda\Lambda$ source will produce low-lying states in the lattice
volume that are predominately linear combinations of the
$\Lambda\Lambda$, $\Sigma^{\pm,0}\Sigma^{\mp,0}$, and $N\Xi$
two-baryon states.

\subsection{Nucleon-nucleon interactions}
\label{sec:nucl-nucl-inter}
\noindent
Perhaps the most studied and best understood of the two-hadron systems
are the proton-proton and proton-neutron.  At low energies, only two
combinations of spin and isospin are possible, a spin-triplet
isosinglet $np\ (^3S_1)$ and a spin-singlet isotriplet $pp\ (^1S_0)$.
At the physical pion mass, the scattering lengths in these channels
are unnaturally large and the $^3S_1$ channel contains a shallow bound
state, the deuteron, with a binding energy of $\sim 2.22~{\rm
  MeV}$. These large scattering lengths and the shallow bound state
arise because the coefficient of the momentum-independent four-nucleon
operator in the low-energy effective field theory has a non-trivial
ultraviolet fixed-point for the physical light-quark masses.  An
interesting line of investigation is the study of the scattering
lengths as a function of the quark masses to ascertain the sensitivity
of this fine-tuning to the QCD
parameters~\cite{Beane:2002xf,Beane:2002vs,Epelbaum:2002gb}.  The fine
tuning is not expected to persist away from the physical masses and we
expect our present (unphysical) calculations to yield scattering
lengths that are natural-sized.

\begin{figure}[!th]
  \centering
  \includegraphics[width=1.0\columnwidth]{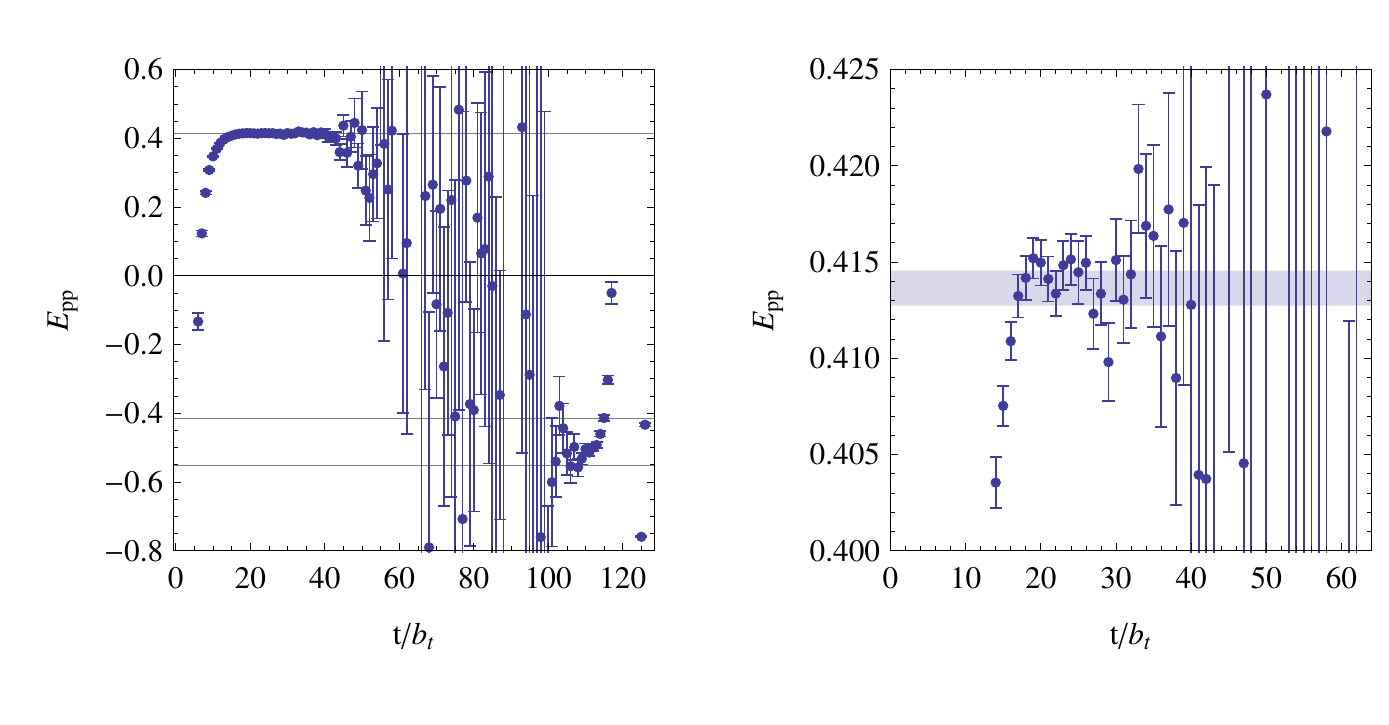}
  \caption{The left panel is the proton-proton $(^1S_0)$ GEMP with
    $t_J=1$, while the right panel shows the plateau region of the
    left panel.  The band in the right panel and the upper line in the
    left panel correspond to $2 M_N$, while the lower two lines in the
    left panel correspond to $-2 M_N$ and $-2 (M_N + m_\pi)$,
    respectively.  }
  \label{fig:pp-BOT}
\end{figure}
\begin{figure}[!th]
  \centering
  \includegraphics[width=1.0\columnwidth]{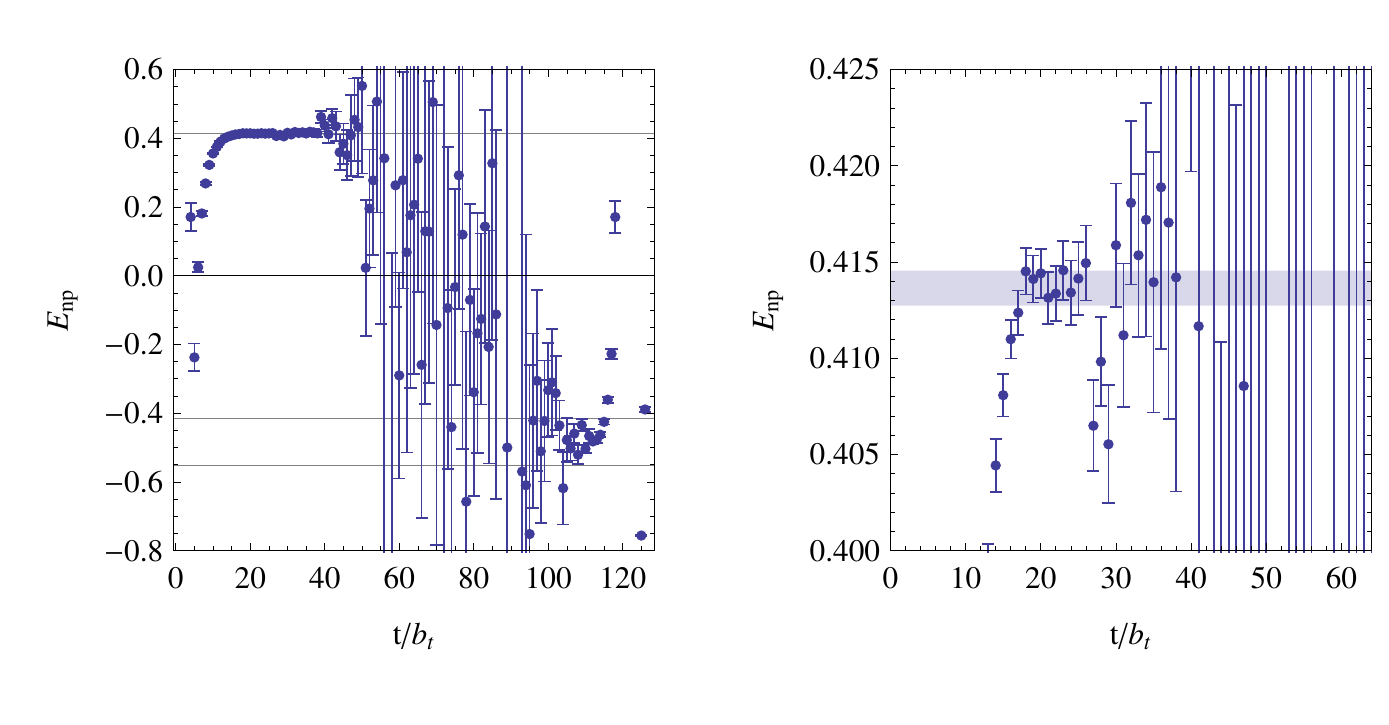}
  \caption{The left panel is the neutron-proton $(^3S_1)$ GEMP with
    $t_J=1$, while the right panel shows the plateau region of the
    left panel.  The band in the right panel and the upper line in the
    left panel correspond to $2 M_N$, while the lower two lines in the
    left panel correspond to $-2 M_N$ and $-2 (M_N + m_\pi)$,
    respectively.  }
  \label{fig:np-BOT}
\end{figure}
\begin{figure}[!th]
  \centering
  \includegraphics[width=1.0\columnwidth]{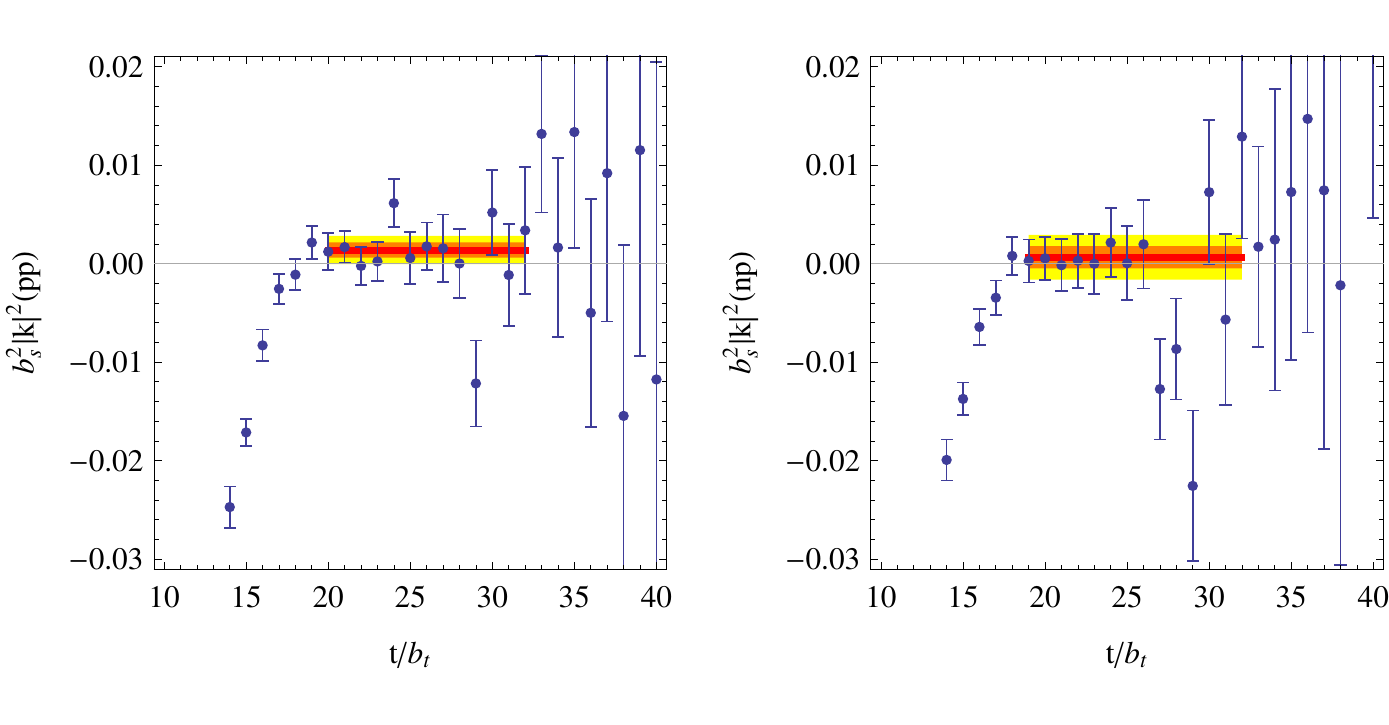}
  \caption{The left panel is the effective $|{\bf k}|^2$ plot for the
    proton-proton $(^1S_0)$ channel with $t_J=1$ and the fit to the
    plateau.  The right panel is for the neutron-proton $(^3S_1)$
    channel. }
  \label{fig:NN-k2-BOT}
\end{figure}
\begin{figure}[!th]
  \centering
  \includegraphics[width=1.0\columnwidth]{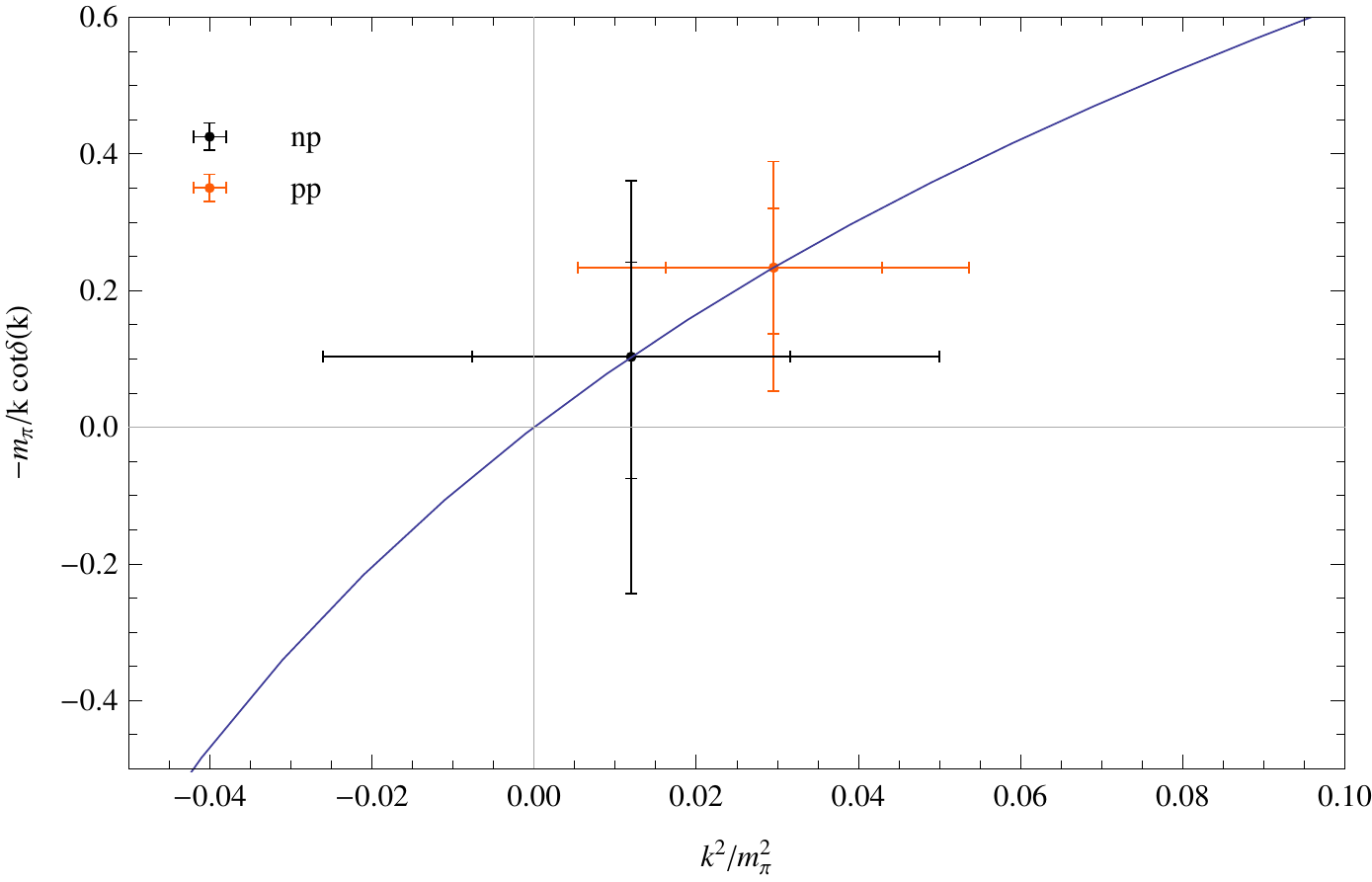}\\
  \caption{The inverse of the real part of the inverse scattering
    amplitude normalized to the pion mass as a function of the
    squared-momentum in the center-of-mass normalized to the pion
    mass.  The dotted curve corresponds to the inverse ``S-function'',
    defined in eq.~(\protect\ref{eq:1}), from which
    $(k\cot\delta)^{-1}$ is determined from $|{\bf k}|^2$.  The inner
    uncertainty of each data point is statistical and the outer
    uncertainty is the statistical and systematic uncertainty combined
    in quadrature.  }
  \label{fig:NN-INVpcotVk2}
\end{figure}

The GEMP obtained from the proton-proton correlation functions is
shown in fig.~\ref{fig:pp-BOT}, and fig.~\ref{fig:np-BOT} shows the
analogous plots for the neutron-proton correlation function.
After the initial plateau region, the GEMPs show a slight downward
fluctuation at time-slice $\sim 29$, which we believe is statistical
in nature.\footnote{This fluctuation appears in both the proton-proton
  and neutron-proton GEMPs and in several of the other channels.  The
  fact that the feature appears in multiple correlation functions is
  not a surprise as all the correlation functions are generated from
  the same light-quark and strange-quark propagators.}
Figure~\ref{fig:NN-k2-BOT} shows the effective $|{\bf k}|^2$
plot~\footnote{ For the presentation of the results of the calculation
  we use $|{\bf k}|^2$ to denote $q_n^2$, as the GEMPs do not isolate
  a particular energy-eigenvalue or eigenstate.} for both the
proton-proton and neutron-proton channels.
Both channels exhibit plateaus in $|{\bf k}|^2$.  The plateau in the
neutron-proton channel is consistent with zero, while the plateau in
the proton-proton channel differs from zero at the $\sim
1\sigma$-level.  We conclude that at this value of the pion mass, the
interactions between nucleons produce a small scattering length in
both channels compared to the naive estimate of $m_\pi^{-1}\sim
0.5~{\rm fm}$.  The results of the analysis are shown in
Table~\ref{tab:NNresults}. Extracted lattice quantities are converted
to physical units using $b_s=0.1227(8)$ fm; here the lattice spacing
uncertainty is sub-dominant.
\begin{table}[!ht]
  \caption{Results for the pp $(^1S_0)$ and np $(^3S_1)$ channels.}
  \label{tab:NNresults}
  \begin{ruledtabular}
    \begin{tabular}{ccccccc}
      Process       
      &  $|{\bf k}|^2/m_\pi^2$ 
      &  $\Delta E$ (MeV) 
      &  $-1/p\cot\delta$ (fm)
      &  $\chi^2/{\rm dof}$ 
      &  fitting interval
      \\
      \hline 
      & \\
      $pp$ 
      & $0.030(13)(20)$ 
      & $3.9(1.7)(2.6)$ 
      & ${0.118^{+0.044}_{-0.049}}^{+0.065}_{-0.077}$ 
      & $ 1.6 $ & $20\rightarrow 32$
      \\
      & \\
      $np$ 
      & $0.012(20)(33)$ 
      & $1.6(2.6)(4.3)$ &
      ${0.052^{+0.07}_{-0.09}}^{+0.11}_{-0.15}$ 
      & $1.96$ & $19\rightarrow 32$ \\
      & 
    \end{tabular}
  \end{ruledtabular}
\end{table}
Motivated by the fact that the pion mass dictates the range of the
interaction between nucleons, we have shown $m_\pi/p\cot\delta(p)$ as
a function of $|{\bf k}|^2/m_\pi^2$ in fig.~\ref{fig:NN-INVpcotVk2}.

\begin{figure}[!th]
  \centering
  \includegraphics[width=0.475\columnwidth]{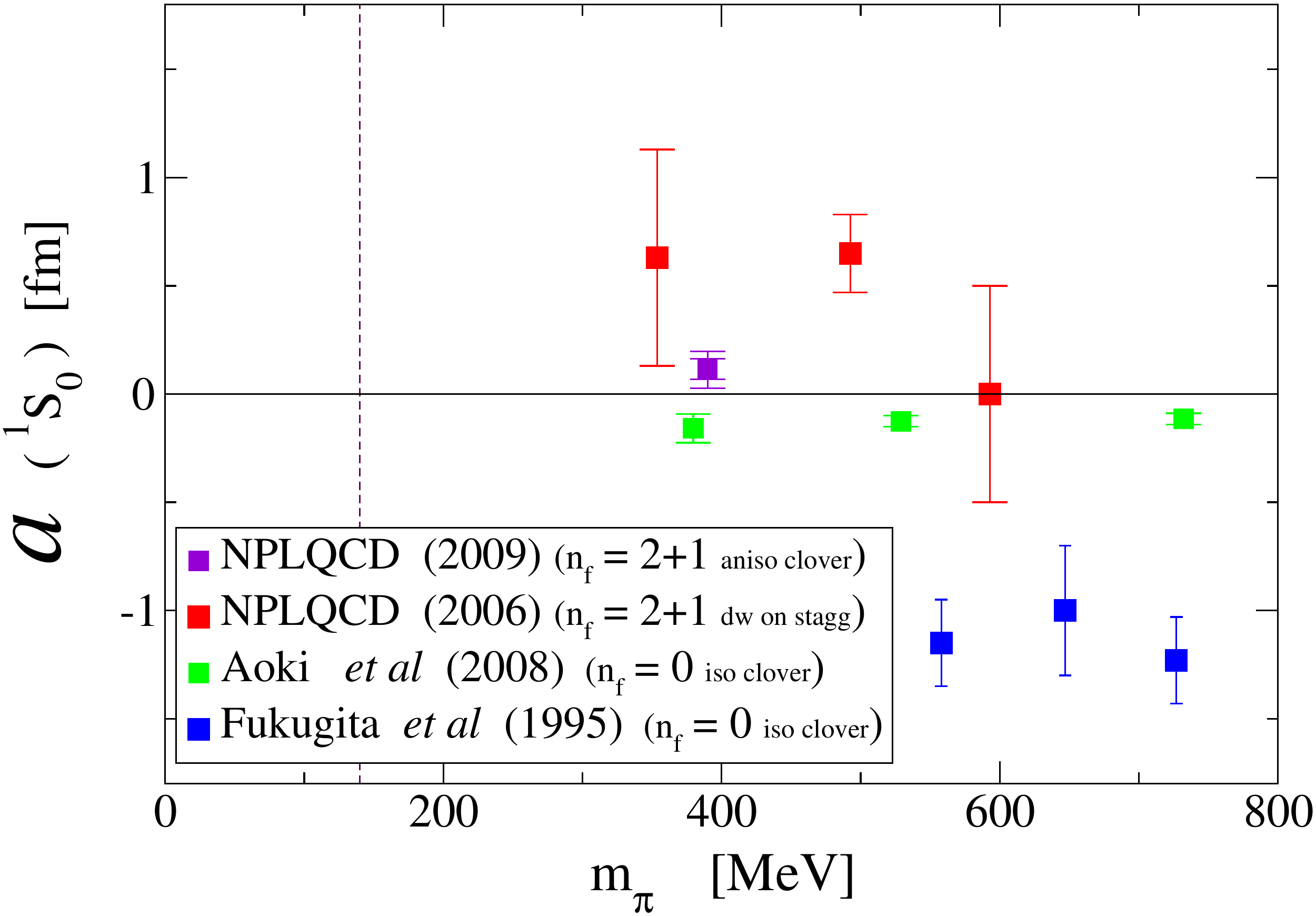}\
  \ \ \
  \includegraphics[width=0.475\columnwidth]{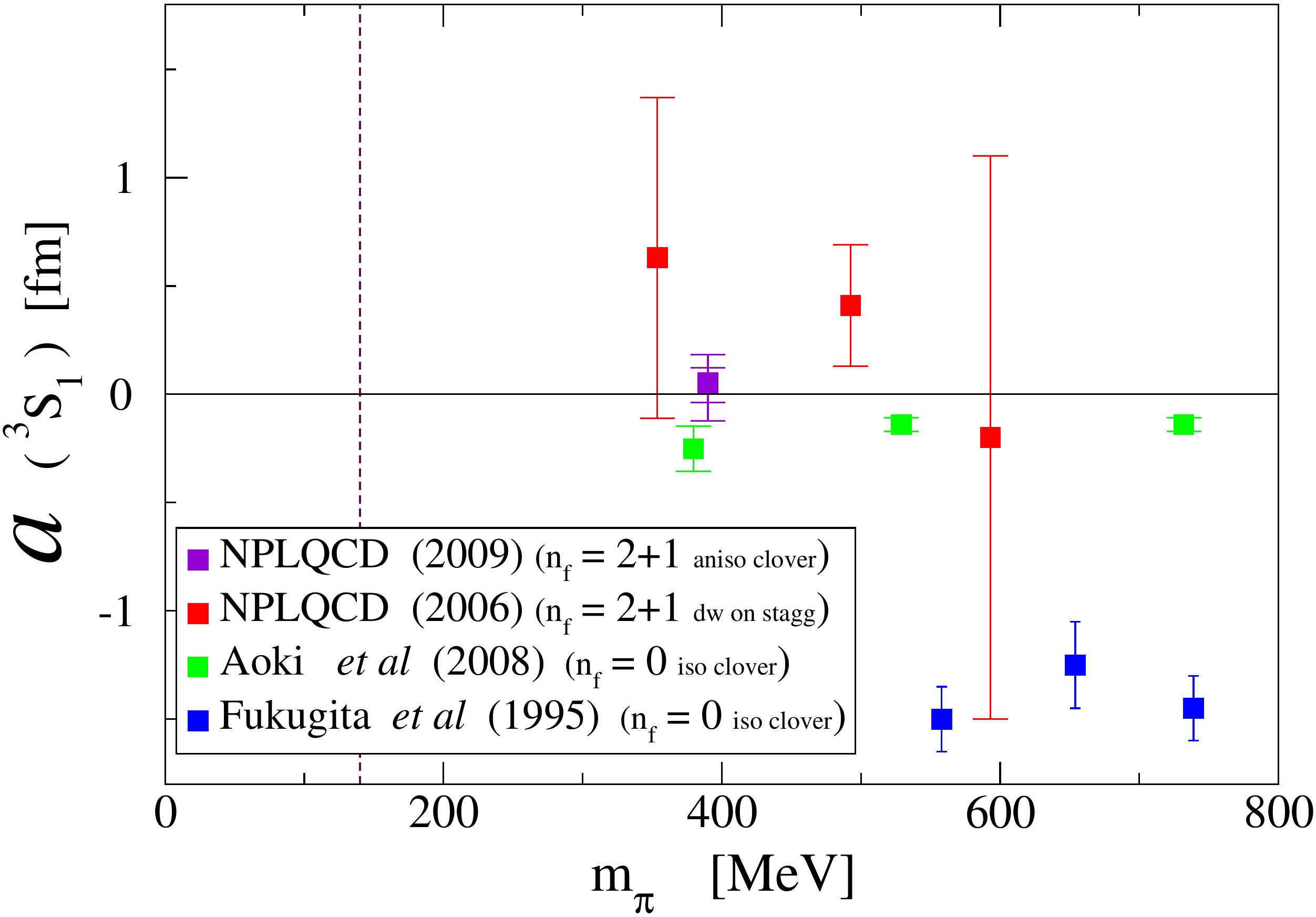}\\
  \caption{A compilation of the scattering lengths for NN scattering
    in the $^1S_0$ (left panel) and $^3S_1$ (right panel) calculated
    in lattice QCD and quenched lattice QCD. The data are from
    Refs.~\protect{\cite{Fukugita:1994ve}}, \protect{\cite{Beane:2006mx}},
      \protect{\cite{Aoki:2008hh}} and the current work. The vertical
      dashed line corresponds to the physical pion mass. }
  \label{fig:NN-ALL-LQCD}
\end{figure}

A summary of lattice QCD calculations of NN scattering is shown in
fig.~\ref{fig:NN-ALL-LQCD}. Since the momenta at which the phase
shifts are measured are small, we present these results as scattering
lengths, implicitly assuming that higher order coefficients in the
effective range expansion, eq.~(\ref{eq:ere}), are natural sized. The
results calculated in this work are consistent with those that we
obtained using mixed-action lattice QCD~\cite{Beane:2006mx}.  It is
interesting to note that the results of quenched
calculations~\cite{Aoki:2008hh} yield scattering lengths that are
consistent within uncertainties with the fully-dynamical $n_f=2+1$
values.
%

\subsection{Hyperon-nucleon interactions ($s=-1$)}
\label{sec:hyper-nucl-inter}
\noindent
Lattice QCD calculations of the YN interactions are of greater
phenomenological importance than those of NN interactions because of
the limited experimental access to hyperon systems in the laboratory
and the possible role of hyperons in nuclear astrophysics.  We have
calculated the energy-eigenvalues of systems with the quantum numbers
of $n\Lambda$ and $n\Sigma^-$ in both spin channels.

\subsubsection{$n\Sigma^-$ interactions  ($I={3\over 2}$) }
\label{sec:NSig}

Interpolating operators with the quantum numbers of $n\Sigma^-$ in
either the $^1S_0$ or $^3S_1$ channels will couple to energy
eigenstates in the lattice volume with strangeness $ s= -1$ and isospin
of $I={3\over 2}$, and the spectrum is not expected to have more than
one low-lying ground state.  We expect to be able to describe these
systems with a single elastic-scattering channel for the lattice
volumes we are working in as these are the only two-baryon states
comprised of octet baryons that have these quantum numbers.

\begin{figure}[!th]
  \centering
  \includegraphics[width=1.0\columnwidth]{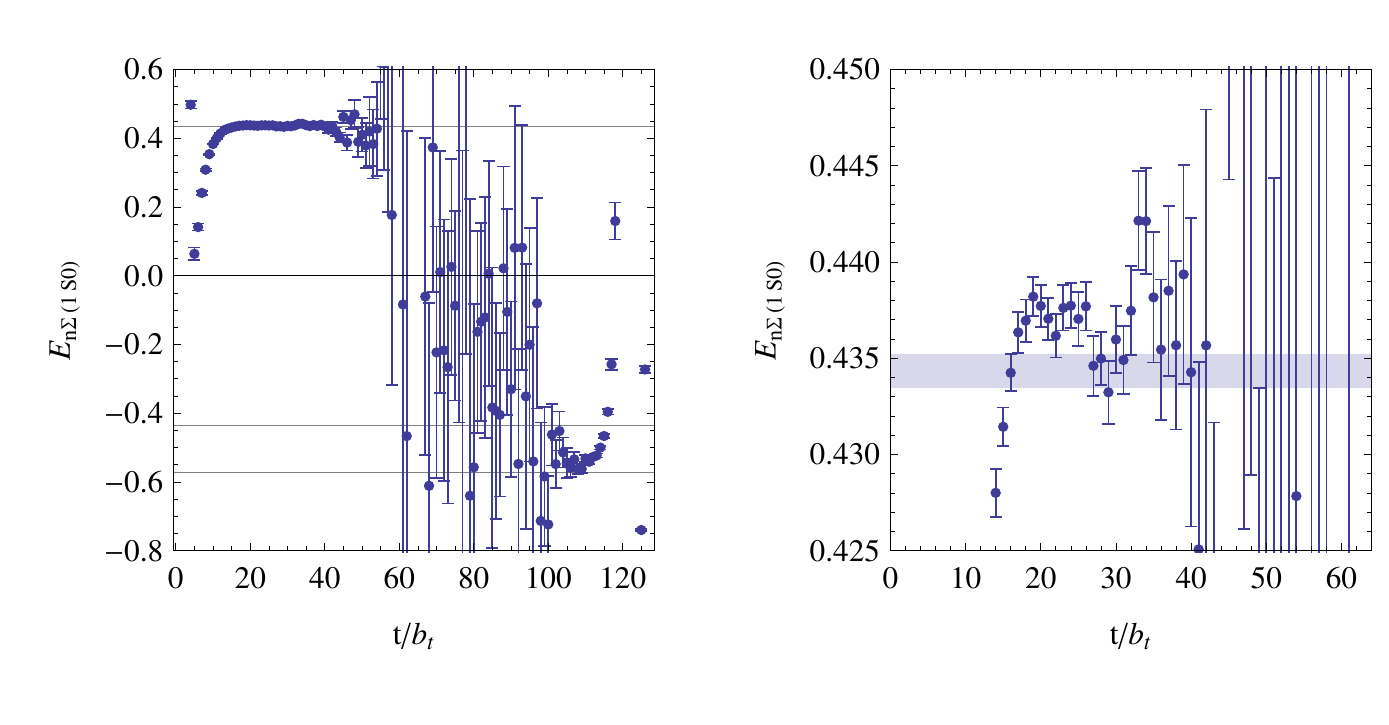}
  \caption{The left panel is the $n\Sigma^-$ $(^1S_0)$ GEMP with
    $t_J=1$, while the right panel shows the plateau region of the
    left panel.  The band in the right panel and the upper line in the
    left panel correspond to $M_\Sigma + M_N$, while the lower two
    lines in the left panel correspond to $-(M_\Sigma + M_N)$ and
    $-(M_\Sigma + M_N + 2 m_\pi)$, respectively.  }
  \label{fig:nSig1s0-BOT}
\end{figure}

\begin{figure}[!th]
  \centering
  \includegraphics[width=1.0\columnwidth]{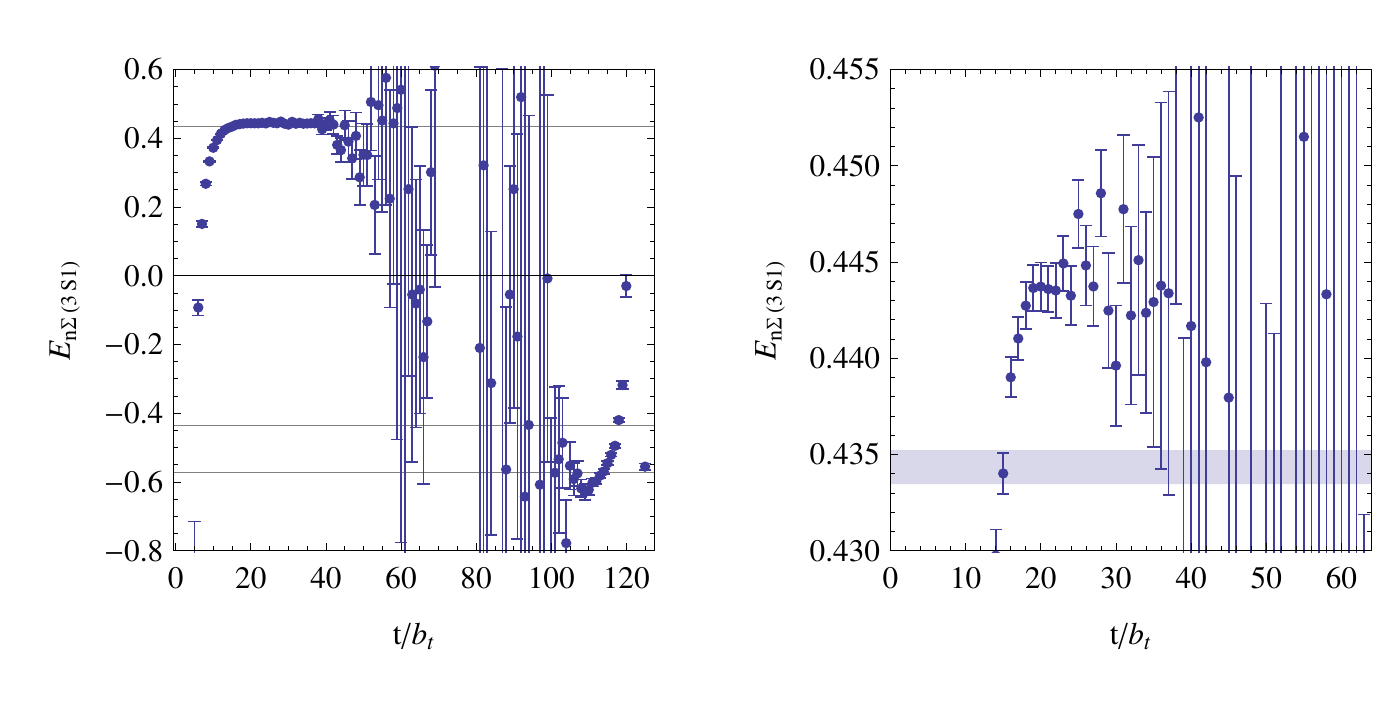}
  \caption{The left panel is the $n\Sigma^-$ $(^3S_1)$ GEMP with
    $t_J=1$, while the right panel shows the plateau region of the
    left panel.  The band in the right panel and the upper line in the
    left panel correspond to $M_\Sigma + M_N$, while the lower two
    lines in the left panel correspond to $-(M_\Sigma + M_N)$ and
    $-(M_\Sigma + M_N + 2 m_\pi)$, respectively.  }
  \label{fig:nSig3s1-BOT}
\end{figure}

\begin{figure}[!th]
  \centering
  \includegraphics[width=1.0\columnwidth]{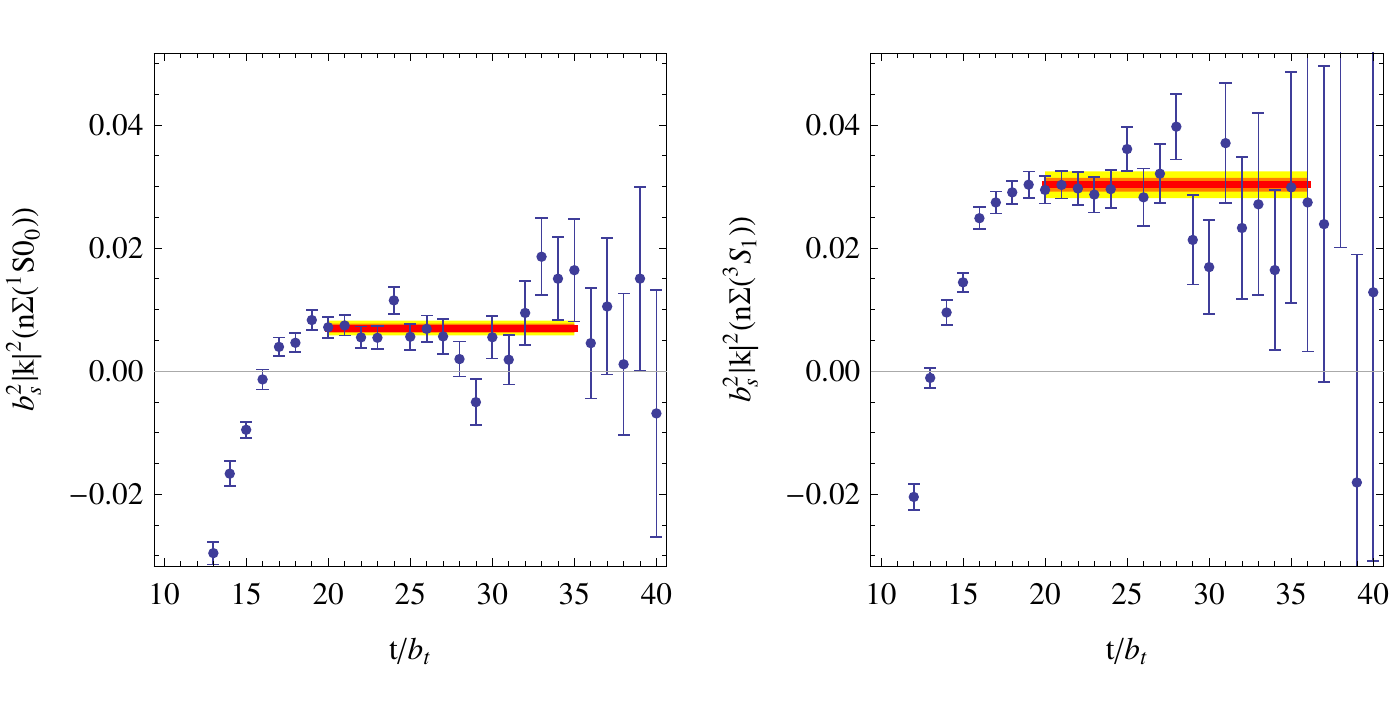}
  \caption{The left panel is the effective $|{\bf k}|^2$ plot for the
    $n\Sigma^-$ $(^1S_0)$ channel with $t_J=1$ and the fit to the
    plateau.  The right panel is for the $n\Sigma^-$ $(^3S_1)$
    channel. }
  \label{fig:NSig-k2-BOT}
\end{figure}

\begin{figure}[!th]
  \centering
  \includegraphics[width=1.0\columnwidth]{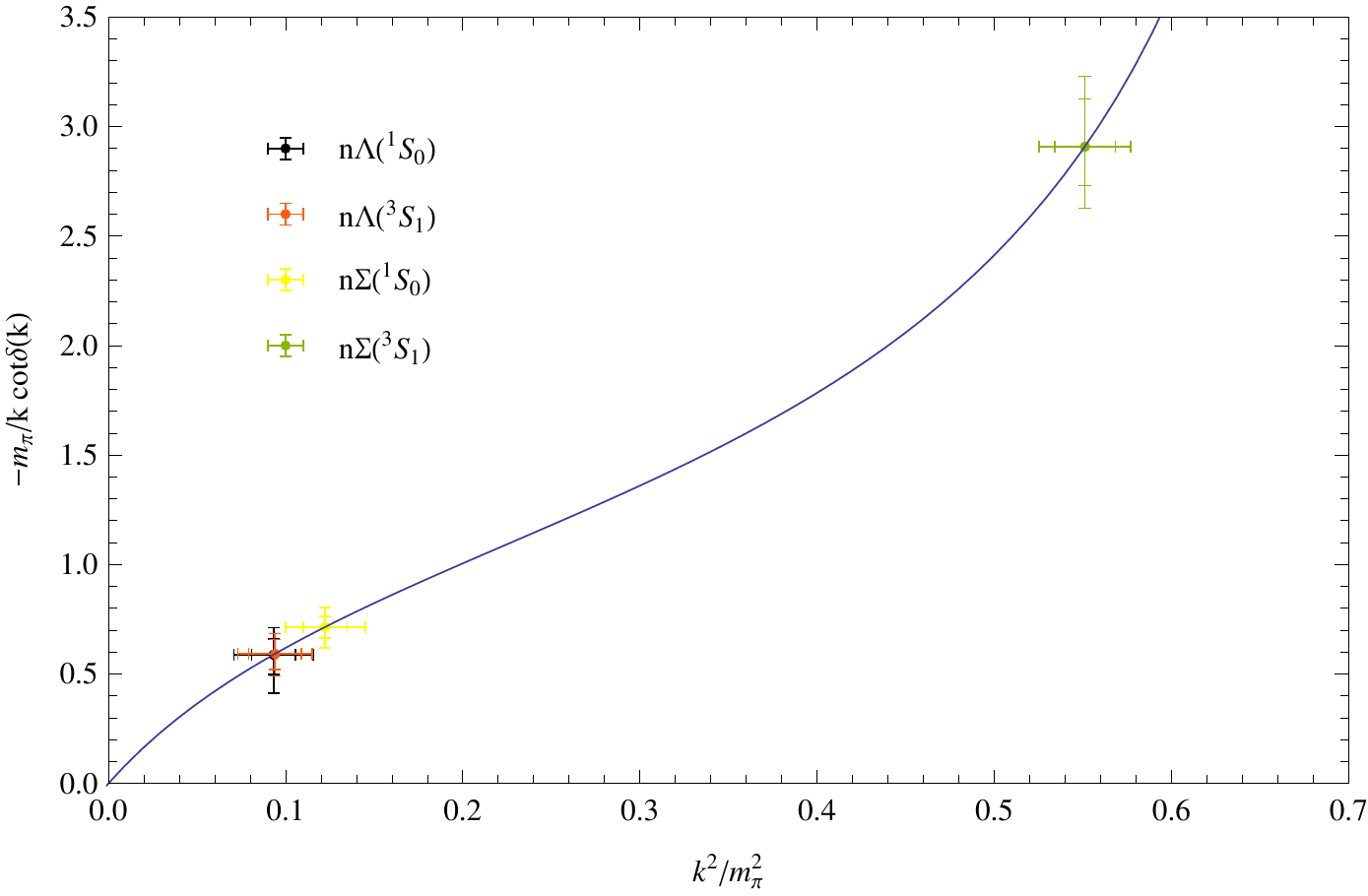}\\
  \caption{The inverse of the real part of the inverse scattering
    amplitude normalized to the pion mass as a function of the
    squared-momentum in the center-of-mass of the baryons normalized
    to the pion mass.  The dotted curve corresponds to the inverse
    ``S-function'', defined in eq.~(\protect\ref{eq:1}), from which
    $(k\cot\delta)^{-1}$ is determined from $|{\bf k}|^2$.  The inner
    uncertainty of each data point is statistical and the outer
    uncertainty is the statistical and systematic uncertainty combined
    in quadrature.  }
  \label{fig:NSig-INVpcotVk2}
\end{figure}

The GEMP for the $n\Sigma^-$ $(^1S_0)$ is shown in
fig.~\ref{fig:nSig1s0-BOT} and exhibits a clear plateau, as does the
GEMP for the $n\Sigma^-$ $(^3S_1)$ channel that is shown in
fig.~\ref{fig:nSig3s1-BOT}.
Figure~\ref{fig:NSig-k2-BOT} shows the effective $|{\bf k}|^2$ plot
for both the $n\Sigma^-$ $(^1S_0)$ and $n\Sigma^-$ $(^3S_1)$ channels.
\begin{table}[!ht]
  \caption{Results for the strangeness $s=-1$ hyperon-nucleon channels.}
  \label{tab:NSigresults}
  \begin{ruledtabular}
    \begin{tabular}{ccccccc}
      Process       
      &  $|{\bf k}|^2/m_\pi^2$
      &  $\Delta E$ (MeV) 
      &  $-1/p\cot\delta$ (fm)
      &  $\chi^2/{\rm dof}$ 
      &  fitting interval
      \\
      \hline 
      & \\
      $n\Sigma^-$ $(^1S_0)$  
      & $0.122(12)(19)$ 
      & $15.3(1.5)(2.3)$ 
      & ${0.361^{+0.025}_{-0.026}}^{+0.038}_{-0.040}$ 
      & $ 2.06 $ & $20\rightarrow 35$
      \\
      & \\
      $n\Sigma^-$ $(^3S_1)$ 
      & $0.551(17)(19)$ 
      & $67.9(2.1)(2.3)$ 
      & ${1.47^{+0.11}_{-0.09}}^{+0.12}_{-0.11}$ 
      & $0.91$ & $19\rightarrow 36$ \\
      \hline 
      & \\
      $n\Lambda$ $(^1S_0)$  
      & $0.093(12)(19)$ 
      & $11.8(1.6)(2.3)$ 
      & ${0.297^{+0.036}_{-0.046}}^{+0.051}_{-0.075}$ 
      & $ 1.69 $ & $20\rightarrow 35$
      \\
      & \\
      $n\Lambda$ $(^3S_1)$ 
      & $0.094(15)(15)$ 
      & $11.9(1.9)(1.9)$ 
      & ${0.299^{+0.033}_{-0.036}}^{+0.033}_{-0.036}$ 
      & $1.05$ & $20\rightarrow 35$ \\
    \end{tabular}
  \end{ruledtabular}
\end{table}
The results of fitting the clear plateaus that are observed in both
channels are shown in Table~\ref{tab:NSigresults}, and
fig.~\ref{fig:NSig-INVpcotVk2} shows the results presented in
Table~\ref{tab:NSigresults} normalized to the pion mass.
The $n\Sigma^-$ $(^1S_0)$ channel is observed to have an interaction
of natural size, and the energy that is measured in the calculation
lies within the regime of applicability of the effective-range
expansion.  In contrast, the interaction in the $n\Sigma^-$ $(^3S_1)$
channel is seen to be large, $|m_\pi/p\cot\delta(p)|\sim3$, and is
well outside the regime of applicability of the effective range
expansion.  At this momenta, the $n\Sigma^-$ $(^3S_1)$ is strongly
interacting in a way that is consistent with an attractive interaction
that supports a bound-state (which we find no direct evidence for in
this calculation), or a repulsive interaction of unnaturally large
range.  These two scenarios cannot be resolved with calculations in a
single volume, but ongoing calculations in different volumes will
resolve this ambiguity.  This result is perhaps the most important
result of this present work.  One concludes from this calculation that
the $n\Sigma^-$ interaction is strongly spin-dependent.

\subsubsection{The coupled $N\Lambda$- $N\Sigma$ channel  ($I={1\over
    2}$) }
\label{sec:NLam}

The strangeness $s=-1$, isospin-${1\over 2}$ energy-eigenstates in the
lattice volume will, in general, couple to both the $N\Lambda$ and
$N\Sigma$ channels.  Therefore, the calculated $N\Lambda$ correlation
functions will receive contributions from all such eigenstates in the
volume.  However, as $M_{\Sigma}-M_{\Lambda} = 0.0051~{\rm t.l.u.}$ is
a significant mass splitting, we expect that a single-channel analysis
is applicable in this system and proceed accordingly. Nevertheless,
calculation of the isospin-${1\over 2}$ $N\Sigma$ correlation function
and the crossed correlation function resulting from a $N\Lambda$
source and a $N\Sigma$ sink would likely improve this analysis.

\begin{figure}[!th]
  \centering
  \includegraphics[width=1.0\columnwidth]{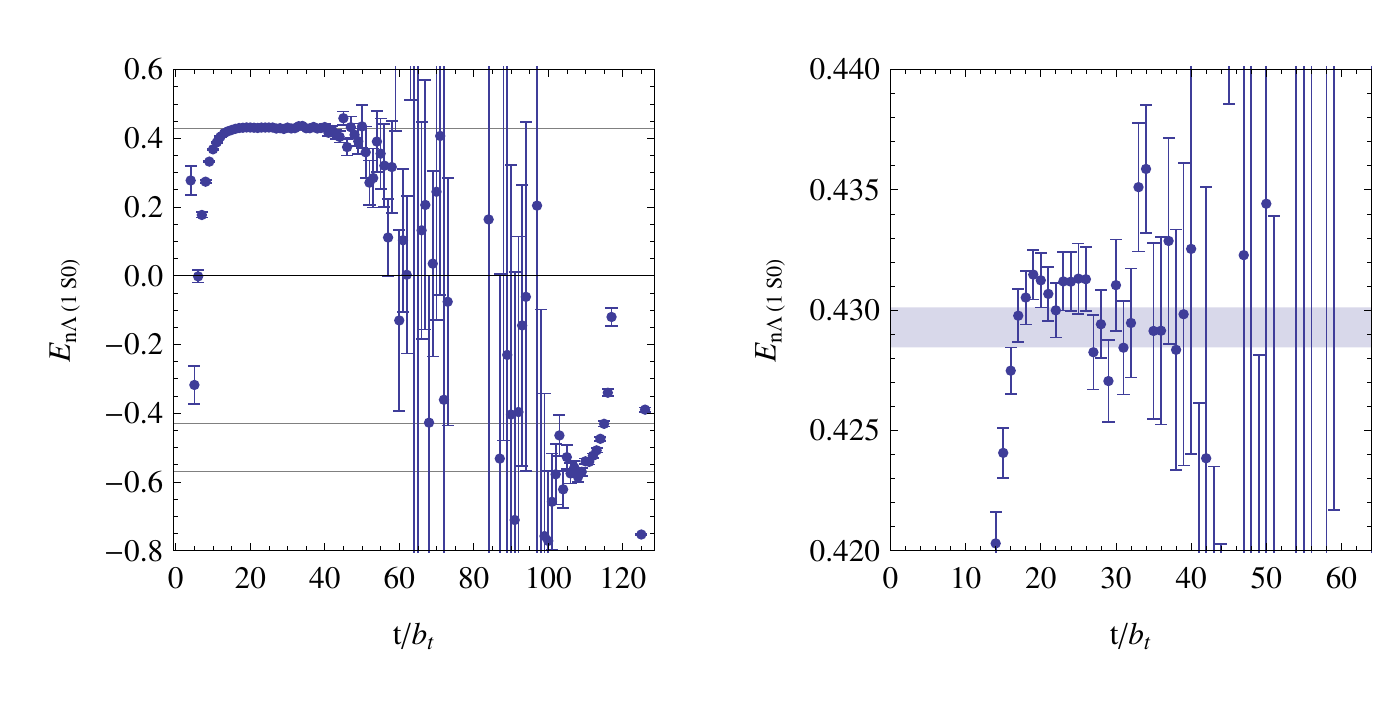}
  \caption{The left panel is the $n\Lambda$ $(^1S_0)$ GEMP with
    $t_J=1$, while the right panel shows the plateau region of the
    left panel.  The band in the right panel and the upper line in the
    left panel correspond to $M_\Lambda + M_N$, while the lower two
    lines in the left panel correspond to $-(M_\Lambda + M_N)$ and
    $-(M_\Lambda + M_N + 2 m_\pi)$, respectively.  }
  \label{fig:nLam1s0-BOT}
\end{figure}

\begin{figure}[!th]
  \centering
  \includegraphics[width=1.0\columnwidth]{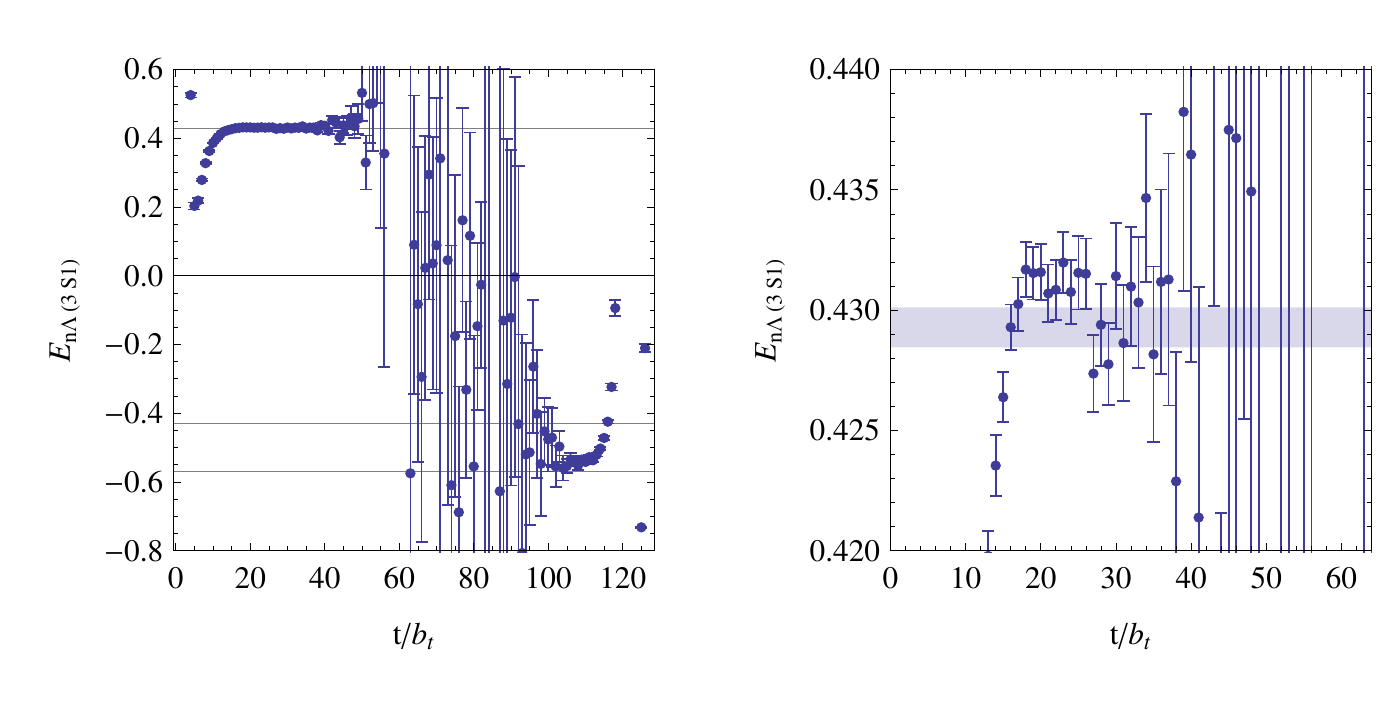}
  \caption{The left panel is the $n\Lambda$ $(^3S_1)$ GEMP with
    $t_J=1$, while the right panel shows the plateau region of the
    left panel.  The band in the right panel and the upper line in the
    left panel correspond to $M_\Lambda + M_N$, while the lower two
    lines in the left panel correspond to $-(M_\Lambda + M_N)$ and
    $-(M_\Lambda + M_N + 2 m_\pi)$, respectively.  }
  \label{fig:nLam3s1-BOT}
\end{figure}

\begin{figure}[!th]
  \centering
  \includegraphics[width=1.0\columnwidth]{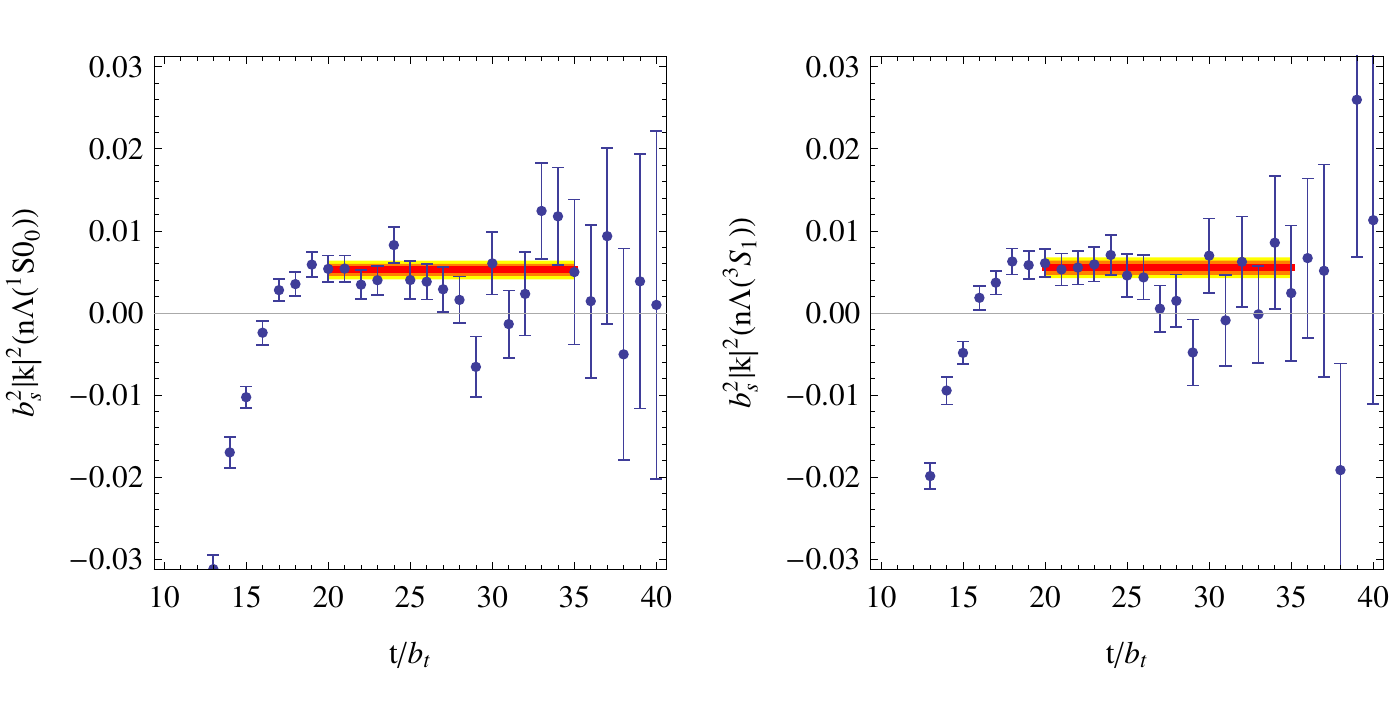}
  \caption{The left panel is the effective $|{\bf k}|^2$ plot for the
    $n\Lambda$ $(^1S_0)$ channel with $t_J=1$ and the fit to the
    plateau.  The right panel is for the $n\Lambda$ $(^3S_1)$
    channel. }
  \label{fig:NLam-k2-BOT}
\end{figure}

The GEMPs for the $n\Lambda$ $(^1S_0)$ and $n\Lambda$ $(^3S_1)$
correlation functions are shown in fig.~\ref{fig:nLam1s0-BOT}
and~\ref{fig:nLam3s1-BOT}, respectively, and in both cases a clear
plateau is visible.
Figure~\ref{fig:NLam-k2-BOT} shows the effective $|{\bf k}|^2$ plot
for both the $n\Lambda$ $(^1S_0)$ and $n\Lambda$ $(^3S_1)$ channels
and clear plateaus are again observed in both channels.
The results of fitting to the plateau region of the effective $|{\bf
  k}|^2$ plots are reported in Table~\ref{tab:NSigresults}, and are
displayed in fig.~\ref{fig:NSig-INVpcotVk2}.  The energy-splittings
are both found to be $\Delta E\sim 0.002~{\rm t.l.u}$, and are
therefore smaller than the expected splitting between the energy
eigenstates resulting from the $N\Lambda$-$N\Sigma$ mixing ($M_\Sigma
- M_\Lambda = 0.0051~{\rm t.l.u.}$).  It therefore seems likely that,
{\it a posteriori}, the single-channel analysis used here is
applicable.  The extracted squared-momenta and hence scattering
amplitudes in the two spin channels are the same within uncertainties.
This indicates that the spin-dependent interactions in these channels
are very small.  If the extracted states are predominately $N\Lambda$,
then this result is expected because of the fact that the long-range
spin-dependent interaction resulting from OPE is absent (as the
$\Lambda$ is an isosinglet).  Further, the extracted squared-momenta
are consistent with that found in the $n\Sigma^-$ $(^1S_0)$ channel.

\subsubsection{Compilation of measurements}
\label{sec:YN-compile}

After the pioneering quenched calculations of YN scattering by
Fukugita {\it et al}~\cite{Fukugita:1994ve}, and the first
fully-dynamical $n_f=2+1$ calculations \cite{Beane:2006gf}, more
refined quenched calculations have been performed along with one
$n_f=2+1$ calculation~\cite{Nemura:2009kc}.  All of the results for
$s=-1$ YN scattering that have been obtained from lattice QCD
calculations are shown in Table~\ref{tab:YNcompile}.
\begin{table}[!t]
  \caption{A compilation of results for strangeness $=-1$ hyperon-nucleon 
    scattering from lattice
    QCD. The columns labeled as ``Valence'' and ``Sea'' list the 
    action used for the valence and sea quarks. A  dash indicates a
    quenched calculation. }
  \label{tab:YNcompile}
  \begin{ruledtabular}
    \begin{tabular}{ccccccc}
      Process 
      &  $m_\pi$ (MeV)      
      &  $|{\bf k}|$ (MeV)  
      &  $-1/p\cot\delta$ (fm)
      &  Valence  
      & Sea 
      &  Reference \\
      \hline 
      $n\Lambda$ $^1S_0$ 
      & $296(3)$
      & $50(26)\ i$
      & $-0.11(7)$
      & Clover & Clover
      & \cite{Nemura:2009kc} \\
      $n\Lambda$ $^1S_0$ 
      & $354(6)$
      & $255(26)$
      & $1.04(28)$
      & DW & Staggered
      & \cite{Beane:2006gf} \\
      $n\Lambda$ $^1S_0$ 
      & $390.4(4.4)$
      & $119(14)$
      & $0.297(76)$
      & Clover & Clover
      & present work \\
      $n\Lambda$ $^1S_0$ 
      & $465(1)$
      & $22(3)\ i$
      & $-0.09(3)$
      & Clover & --
      & \cite{Nemura:2009kc} \\
      $n\Lambda$ $^1S_0$ 
      & $493(8)$
      & $197(24)$
      & $0.63(12)$
      & DW & Staggered
      & \cite{Beane:2006gf} \\
      $n\Lambda$ $^1S_0$ 
      & $514(1)$
      & $17(3)\ i$
      & $-0.07(3)$
      & Clover & --
      & \cite{Nemura:2009kc} \\
      \hline
      $n\Lambda$ $^3S_1$ 
      & $296(3)$
      & $40(24)\ i$
      & $-0.07(7)$
      & Clover & Clover
      & \cite{Nemura:2009kc} \\
      $n\Lambda$ $^3S_1$ 
      & $354(6)$
      & $168(65)$
      & $0.50(27)$
      & DW & Staggered
      & \cite{Beane:2006gf} \\
      $n\Lambda$ $^3S_1$ 
      & $390.4(4.4)$
      & $119(14)$
      & $0.299(49)$
      & Clover & Clover
      & present work \\
      $n\Lambda$ $^3S_1$ 
      & $465(1)$
      & $24(3)\ i$
      & $-0.11(3)$
      & Clover & --
      & \cite{Nemura:2009kc} \\
      $n\Lambda$ $^3S_1$ 
      & $514(1)$
      & $20(2)\ i$
      & $-0.09(2)$
      & Clover & --
      & \cite{Nemura:2009kc} \\
      \hline
      $n\Sigma^-$ $^1S_0$ 
      & $390.4(4.4)$
      & $136(12)$
      & $0.361(46)$
      & Clover & Clover
      & present work \\
      $n\Sigma^-$ $^1S_0$ 
      & $493(8)$
      & $179(30)$
      & $0.57(14)$
      & DW & Staggered
      & \cite{Beane:2006gf}\\
      \hline
      $n\Sigma^-$ $^3S_1$ 
      & $390.4(4.4)$
      & $289.8(6.8)$
      & $1.47(16)$
      & Clover & Clover
      & present work \\
      $n\Sigma^-$ $^3S_1$ 
      & $493(8)$
      & $261(37)$
      & $1.19(53)$
      & DW & Staggered
      & \cite{Beane:2006gf}\\ 
      $n\Sigma^-$ $^3S_1$ 
      & $592(10)$
      & $226(30)$
      & $0.85(22)$
      & DW & Staggered
      & \cite{Beane:2006gf}\\
    \end{tabular}
  \end{ruledtabular}
\end{table}
As the measurements have been performed at different pion masses and
in different lattice volumes, resulting in different center-of-mass
energies, it is difficult to present these results in a single
diagram.  The present measurements and the mixed-action
measurements~\cite{Beane:2006gf} were both on lattices with spatial
volumes of $V\sim (2.5~{\rm fm})^3$, and with a lattice spacing of
$b\sim 0.125~{\rm fm}$ and so a direct comparison is possible for this
subset of measurements.

\subsection{ Hyperon-hyperon interactions ($s=-2$)}
\label{sec:hyper-hyper-inter}
\noindent
Lattice QCD calculations of hyperon-hyperon interactions are important
as they can provide guidance to the experimental programs in
hyper-nuclear physics. They can also improve upon the current understanding
of the stability of the core of supernovae if it becomes energetically
favorable to have strange baryons present.  To this end we have
calculated the correlation functions resulting from $\Lambda\Lambda$
and $\Sigma^-\Sigma^-$ sources and sinks.  The $\Sigma^-\Sigma^-$
channel has $I_z=-2$ and, as such, a single, isolated ground-state is
expected which can be used to determine the $\Sigma^-\Sigma^-$
scattering phase shift with a single-channel analysis.  In contrast,
the correlation function produced by the $\Lambda\Lambda$ source is
expected to exhibit two nearly degenerate states as $2M_\Lambda =
0.44492~{\rm t.l.u.}$ and $M_\Xi+M_N = 0.44783~{\rm t.l.u.}$.
Further, the $\Sigma^+\Sigma^-$ state is likely to be close by, $2
M_\Sigma = 0.45504~{\rm t.l.u.}$.  An operator of the form $\Xi^- p$
would help to resolve the states, but we have not calculated this
correlation function, nor the possible mixed correlation
functions. Further, we have not explored the isospin-1 channels, such
as the $N\Xi-\Lambda\Sigma$ coupled channels, but we would expect to
find very-closely spaced energy eigenstates for the pion mass used in
this calculation, making clean extractions very difficult. In
particular, $M_{\Xi^-} + M_n = 0.449~{\rm t.l.u}$ while $M_\Lambda +
M_{\Sigma^-} = 0.452~{\rm t.l.u}$.  Quenched calculations of the
$p\Xi^0$ scattering length have been presented in
Ref.~\cite{Nemura:2008sp}.

\subsubsection{ $\Lambda\Lambda$ interactions ($I=0$) }
\label{sec:LamLam-inter}

\begin{figure}[!th]
  \centering
  \includegraphics[width=1.0\columnwidth]{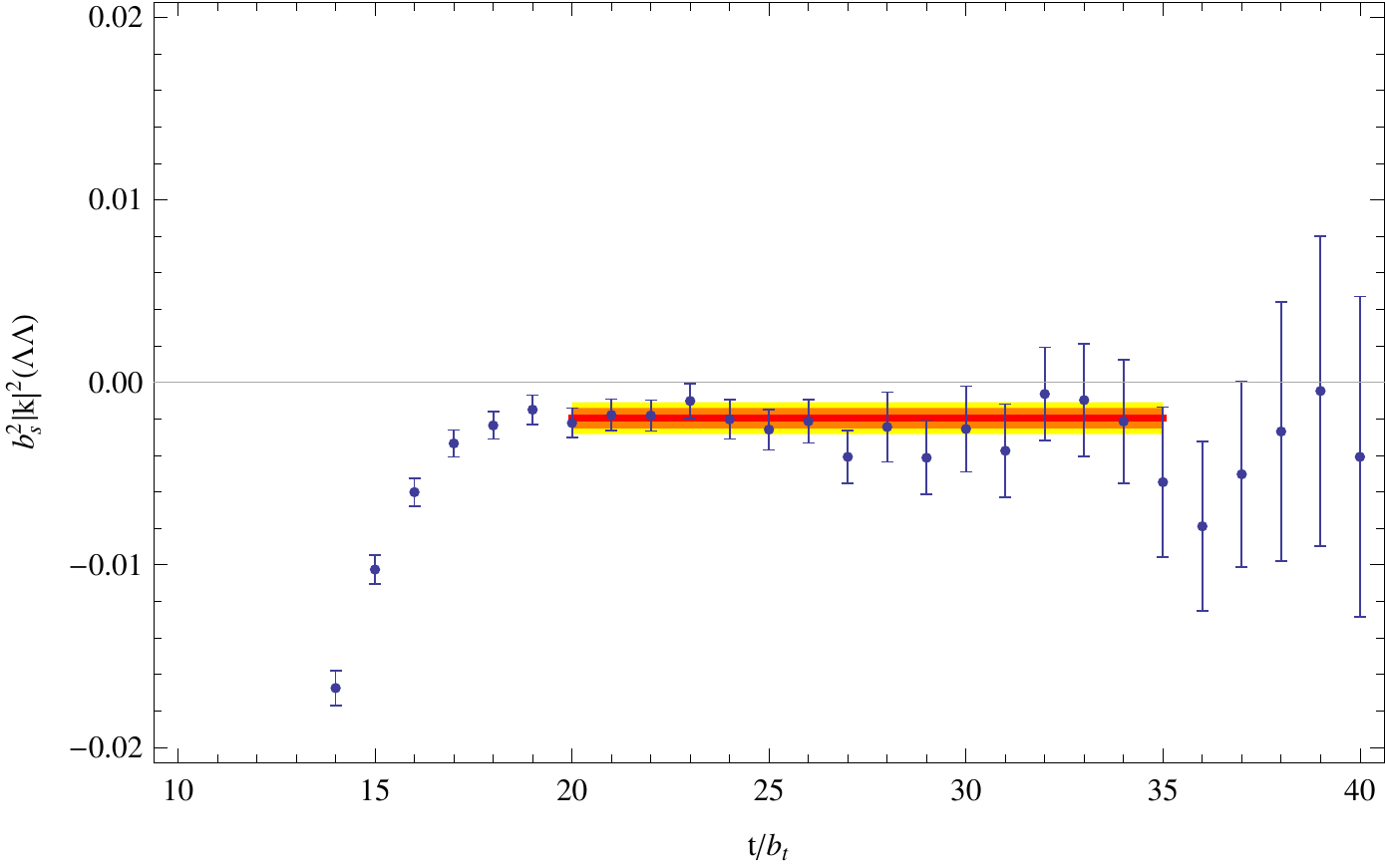}\\
  \caption{The effective $|{\bf k}|^2$ plot for the $\Lambda\Lambda$
    channel with $t_J=3$ and the fit to the plateau.  }
  \label{fig:LamLam-k2-BOT}
\end{figure}

\begin{figure}[!th]
  \centering
  \includegraphics[width=1.0\columnwidth]{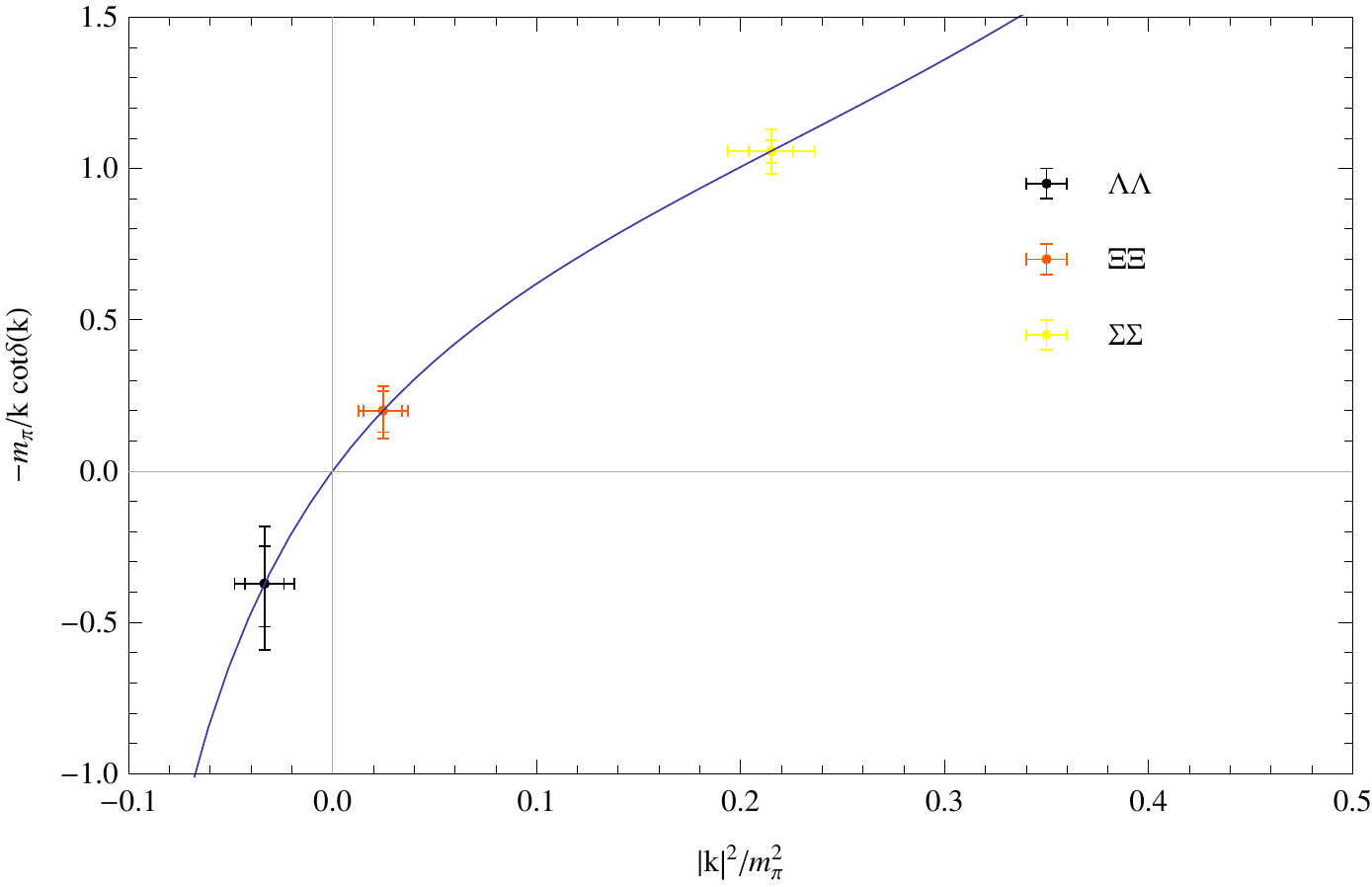}\\
  \caption{The inverse of the real part of the inverse scattering
    amplitude normalized to the pion mass as a function of the
    squared-momentum in the center-of-mass of the baryons normalized
    to the pion mass.  The dotted curve corresponds to the inverse
    ``S-function'', defined in eq.~(\protect\ref{eq:1}), from which
    $(k\cot\delta)^{-1}$ is determined from $|{\bf k}|^2$.  The inner
    uncertainty of each data point is statistical and the outer
    uncertainty is the statistical and systematic uncertainty combined
    in quadrature.  }
  \label{fig:LamLam-INVpcotVk2}
\end{figure}

The low-lying eigenstates that couple to a $\Lambda\Lambda$ source and
sink are, in principle, linear combinations of all two-baryon states
with $s=-2$ and $I=0$, namely, $\Lambda\Lambda$, $\Sigma\Sigma$, $N\Xi$
and their excitations.
\begin{figure}[!th]
  \centering
  \includegraphics[width=1.0\columnwidth]{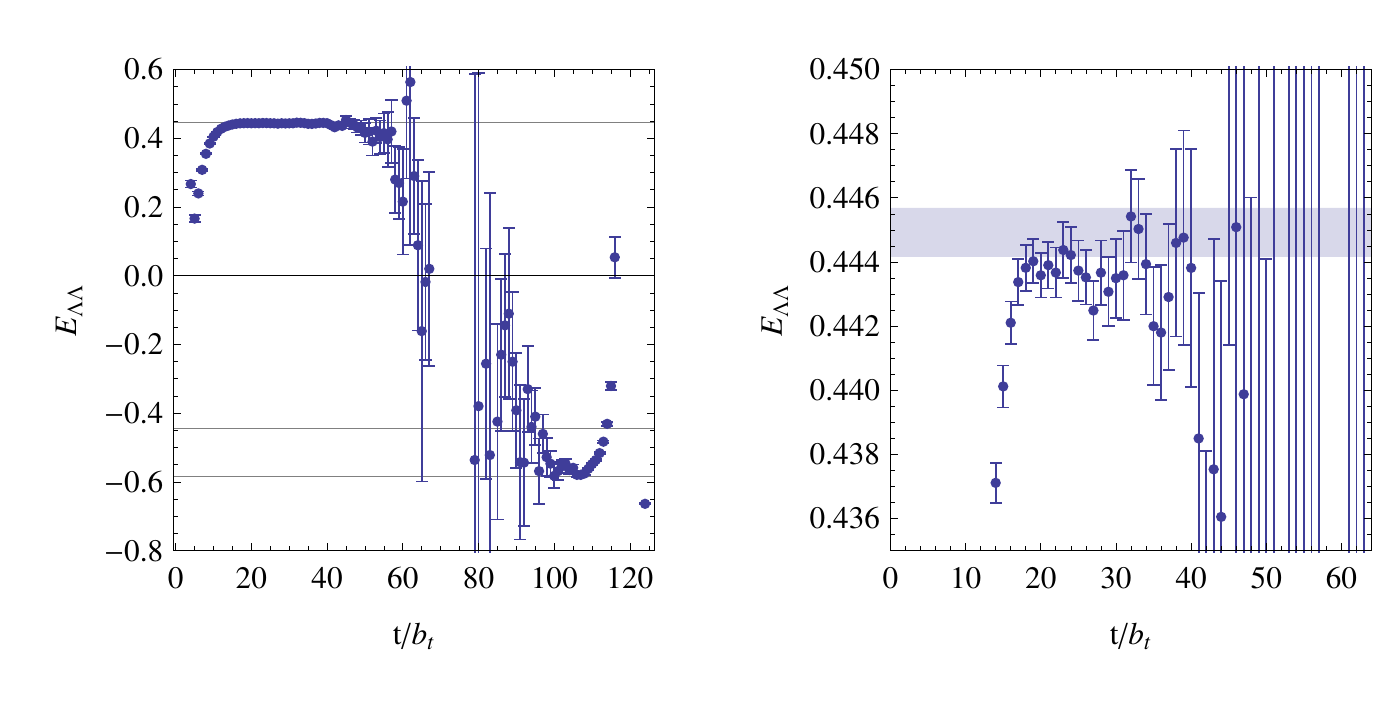}
  \caption{The left panel is the $\Lambda\Lambda$ GEMP with $t_J=3$,
    while the right panel shows the plateau region of the left panel.
    The band in the right panel and the upper line in the left panel
    correspond to $2 M_\Lambda$, while the lower two lines in the left
    panel correspond to $-2 M_\Lambda $ and $-2 (M_\Lambda + m_\pi)$,
    respectively.  }
  \label{fig:LamLam-BOT}
\end{figure}
The GEMP for the $\Lambda\Lambda$ channel (for a particular
combination of SP and SS correlation functions) is shown in
fig.~\ref{fig:LamLam-BOT}, in which a clear plateau is observed.  The
effective $|{\bf k}|^2$ plot shown in fig.~\ref{fig:LamLam-k2-BOT}
also shows a clear plateau that is negative shifted in energy,
unambiguously indicating an attractive interaction.
The energy of this state is the lowest, and the plateau is the
longest, for any combination of SS and SP correlation functions that
could be constructed.  Other combinations of correlation functions
suggest states at higher energies, consistent with expectations of
nearby states, however, with the current data, no definitive
statements can be made.  The splittings between the asymptotic states,
$\Lambda\Lambda$ and $N\Xi$ are such that at measured (very small)
$\Lambda\Lambda$ center-of-mass momentum a single-channel analysis can
be performed to define the $\Lambda\Lambda$ elastic scattering
amplitude, the results of which are shown in
Table~\ref{tab:LamLamresults},
\begin{table}[!ht]
  \caption{Results for the strangeness $=-2$ and strangeness $=-4$ 
    hyperon-hyperon channels.}
  \label{tab:LamLamresults}
  \begin{ruledtabular}
    \begin{tabular}{ccccccc}
      Process       
      &  $|{\bf k}|^2/m_\pi^2$
      &  $\Delta E$ (MeV) 
      &  $-1/p\cot\delta$ (fm)
      &  $\chi^2/{\rm dof}$ 
      &  fitting interval
      \\
      \hline 
      & \\
      $\Lambda\Lambda$  
      & $-0.033(09)(11)$ 
      & $-4.1(1.2)(1.4)$ 
      & ${-0.188^{+0.062}_{-0.072}}^{+0.072}_{-0.085}$ 
      & $ 1.38 $ & $20\rightarrow 35$
      \\
      & \\ 
      \hline 
      & \\
      $\Sigma^-\Sigma^-$
      & $0.215(11)(18)$ 
      & $25.5(1.3)(2.1)$ 
      & ${0.534^{+0.019}_{-0.019}}^{+0.032}_{-0.032}$ 
      & $ 2.6 $ & $20\rightarrow 36$
      \\
      &
      \\
      \hline 
      & \\
      $\Xi^-\Xi^-$
      & $0.0247(94)(77)$ 
      & $2.8(1.1)(0.9)$ 
      & ${0.101^{+0.032}_{-0.036}}^{+0.026}_{-0.029}$ 
      & $ 1.56 $ & $21\rightarrow 34$
      \\
      & 
    \end{tabular}
  \end{ruledtabular}
\end{table}
and shown graphically in fig.~\ref{fig:LamLam-INVpcotVk2}.

The $\Lambda\Lambda$ channel is the only channel in which a
negatively-shifted energy splitting is observed.  However, without
performing measurements on additional lattice volumes, it is presently
not possible to determine if this negative energy shift indicates the
presence of a bound state (the infamous
H-dibaryon~\cite{Jaffe:1976yi}) or if it is simply a continuum state
that is negatively shifted due to an attractive interaction.  The
location of the state on the ``S-function'' curve suggests that it is
fact a continuum state \cite{Beane:2008dv}, but further measurements
are required to properly explore this exciting possibility.

\subsubsection{ $\Sigma^-\Sigma^-$ interactions ($I=2$)}
\label{sec:SigSig-inter}

\begin{figure}[!th]
  \centering
  \includegraphics[width=1.0\columnwidth]{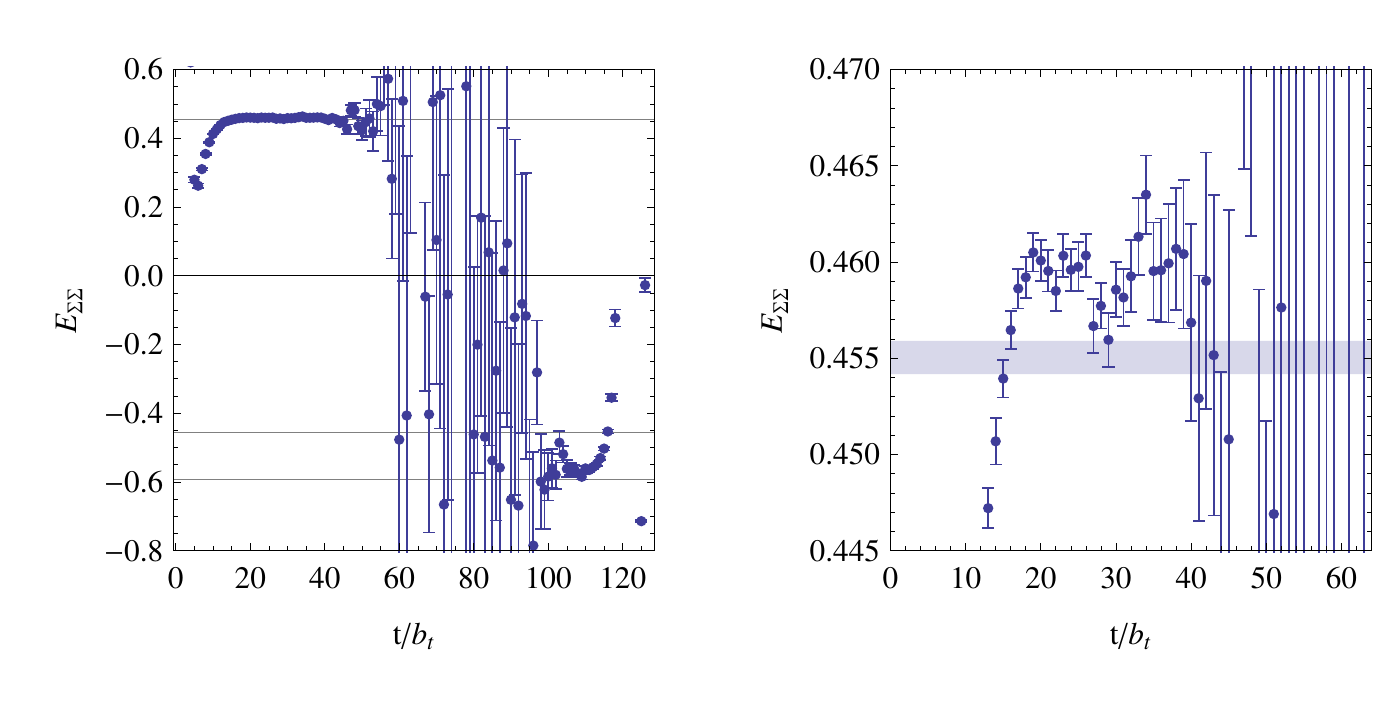}
  \caption{The left panel is the $\Sigma^-\Sigma^-$ GEMP with $t_J=1$,
    while the right panel shows the plateau region of the left panel.
    The band in the right panel and the upper line in the left panel
    correspond to $2 M_\Sigma$, while the lower two lines in the left
    panel correspond to $-2 M_\Sigma $ and $-2 (M_\Sigma + m_\pi)$,
    respectively.  }
  \label{fig:SigSig-BOT}
\end{figure}

\begin{figure}[!th]
  \centering
  \includegraphics[width=1.0\columnwidth]{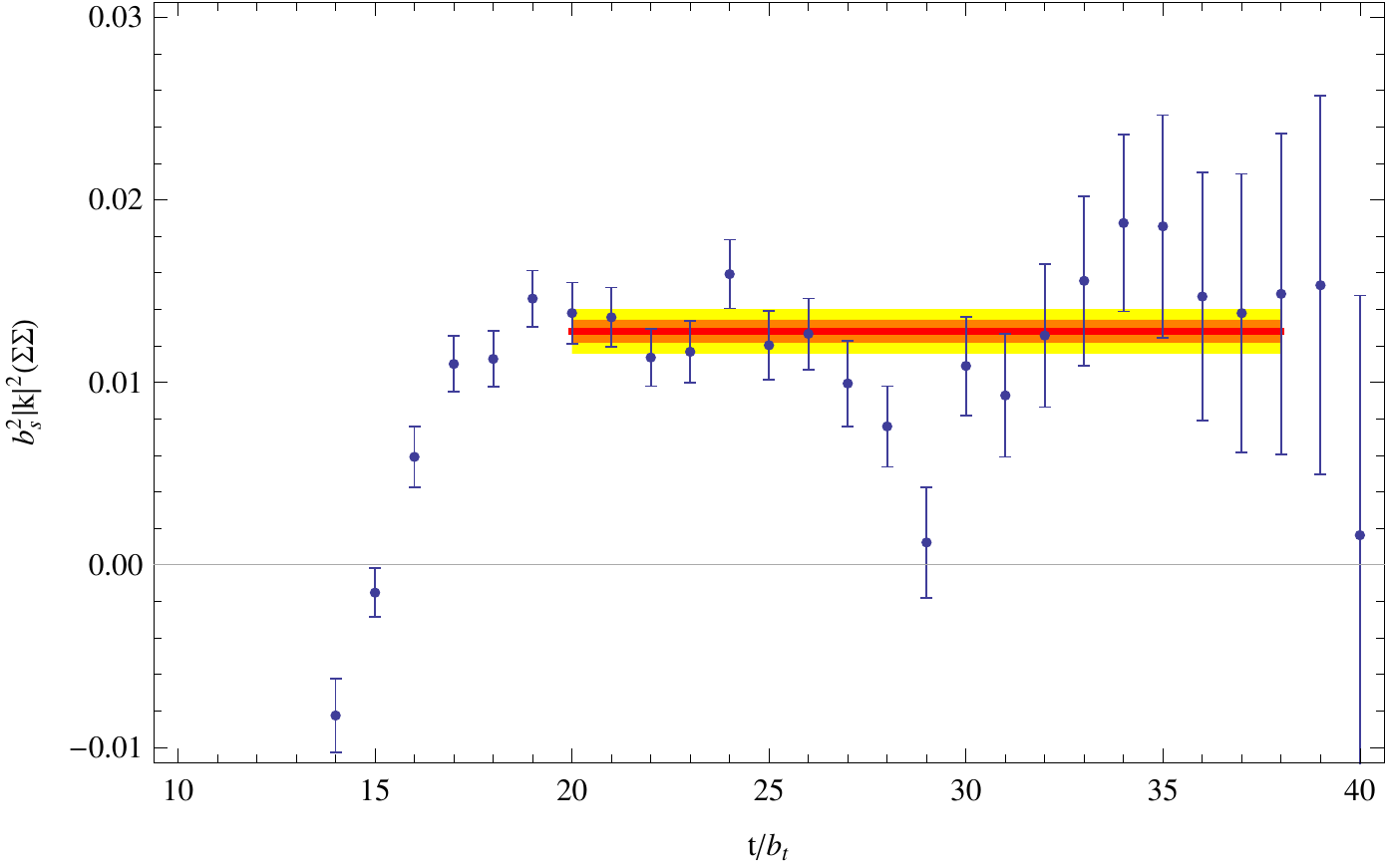}\\
  \caption{The effective $|{\bf k}|^2$ plot for the $\Sigma^-\Sigma^-$
    channel with $t_J=1$ and the fit to the plateau.  }
  \label{fig:SigSig-k2-BOT}
\end{figure}

Due to its quantum numbers, $s=-2$ and $I=2$, the $\Sigma^-\Sigma^-$
channel is not expected to have other states near the ground state in
the lattice volume as there are no other states comprised of two octet
baryons with these quantum numbers.  The GEMP associated with the
$\Sigma^-\Sigma^-$ correlation function is shown in
fig.~\ref{fig:SigSig-BOT}
and the resulting effective $|{\bf k}|^2$ plot is shown in
fig.~\ref{fig:SigSig-k2-BOT}.
The results obtained by fitting to the effective $|{\bf k}|^2$ plot
are shown in Table~\ref{tab:LamLamresults} above, and presented
graphically in fig.~\ref{fig:LamLam-INVpcotVk2}.  The somewhat large
value of $\chi^2/{\rm dof}=2.6$ is due to the downward fluctuation
near time-slice $t=29$ in fig.~\ref{fig:SigSig-k2-BOT}, which we
assume to be statistical in nature.  A smaller fitting interval would
yield approximately the same scattering phase shift, but with a
substantially smaller $\chi^2/{\rm dof}$.

\subsection{ Hyperon-hyperon interactions ($s=-4$)}
\label{sec:hyper-hyper-4-inter}
\noindent
Baryon-baryon interactions in the strangeness -4 sector have no obvious
phenomenological implications.  However, it is found that systems
containing heavier quarks, such as the $\Omega$ or $\Xi$, have
better-behaved correlation functions in lattice QCD calculations.  As
lattice QCD calculations of hadronic interactions are still in their
infancy, it is useful to explore such systems to better understand
aspects of the methodology that we employ.

\begin{figure}[!th]
  \centering
  \includegraphics[width=1.0\columnwidth]{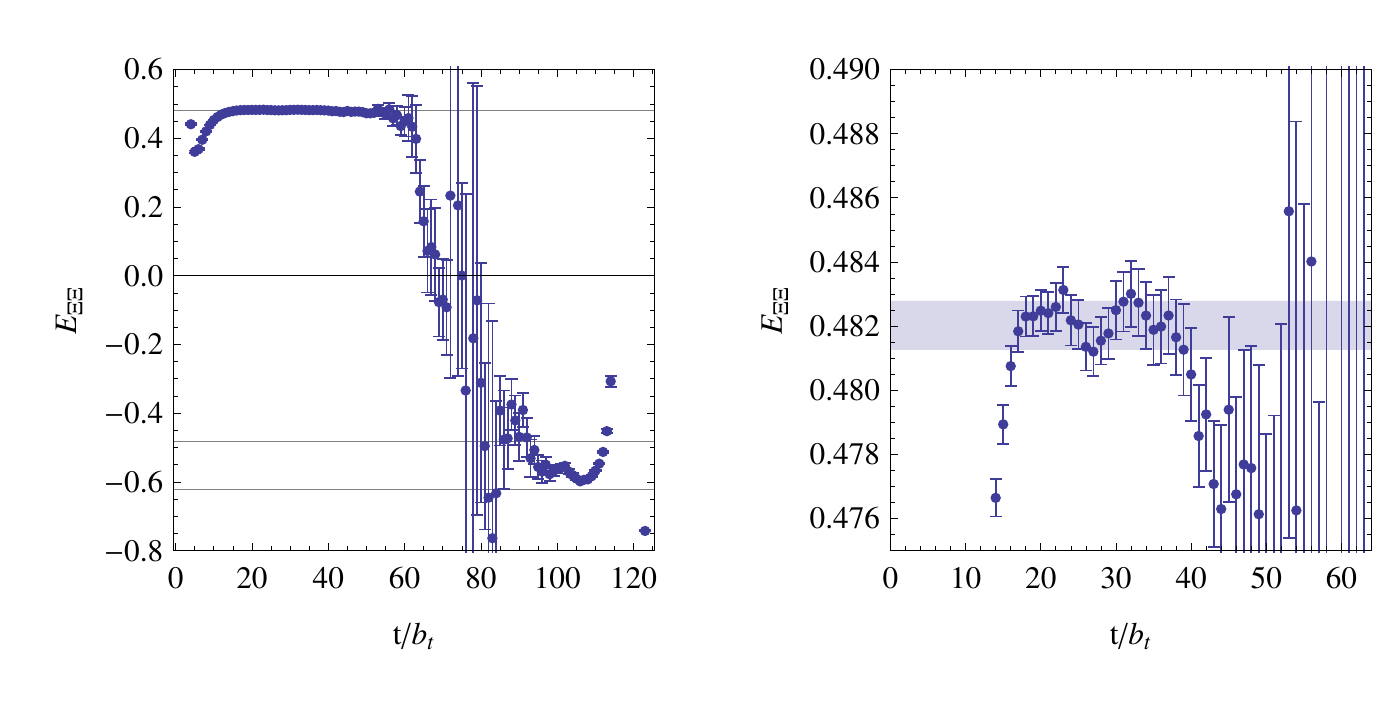}
  \caption{The left panel is the $\Xi^-\Xi^-$ GEMP with $t_J=1$, while
    the right panel shows the plateau region of the left panel.  The
    band in the right panel and the upper line in the left panel
    correspond to $2 M_\Xi$, while the lower two lines in the left
    panel correspond to $-2 M_\Xi $ and $-2 (M_\Xi + m_\pi)$,
    respectively.  }
  \label{fig:XiXi-BOT}
\end{figure}

\begin{figure}[!th]
  \centering
  \includegraphics[width=1.0\columnwidth]{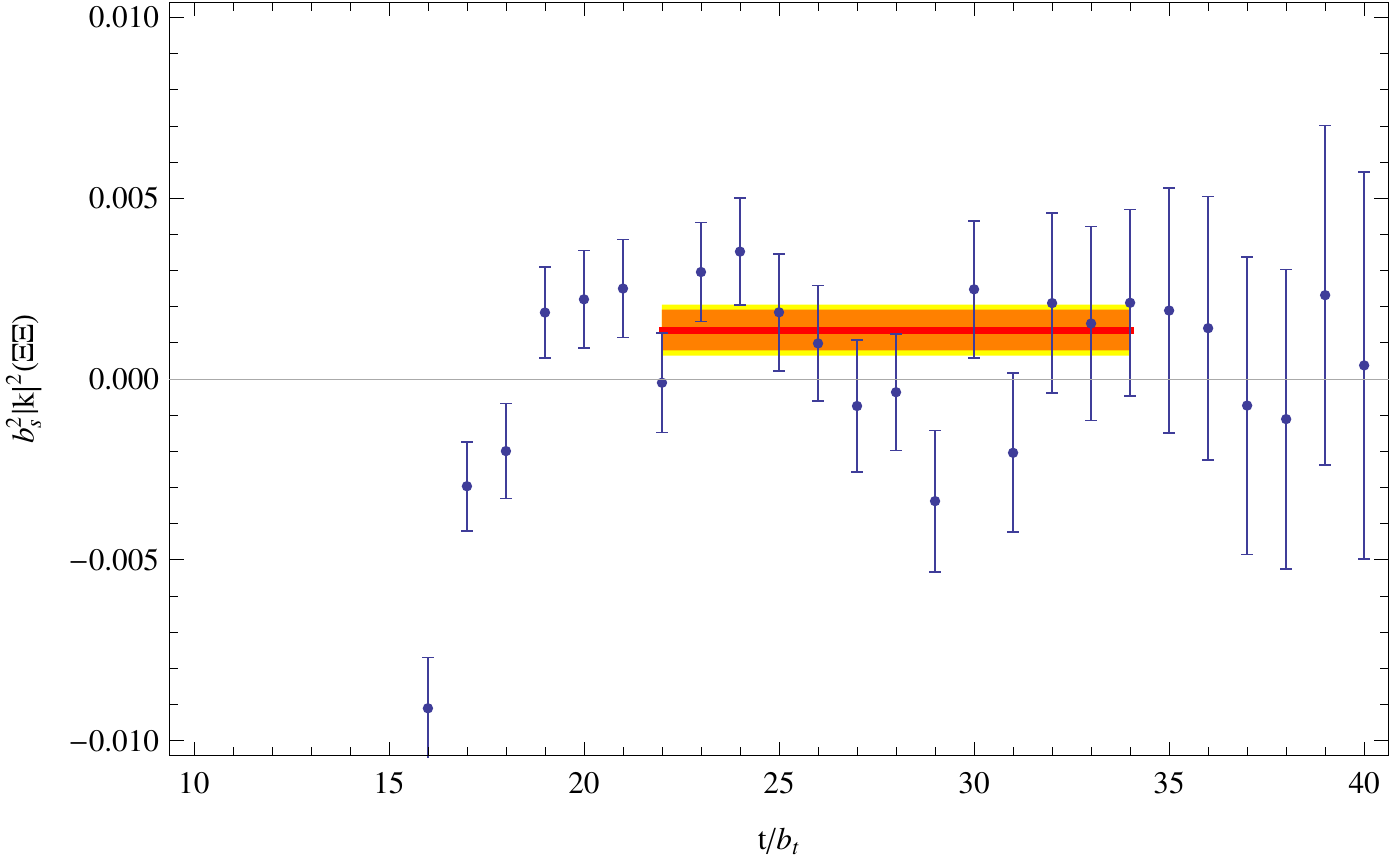}\\
  \caption{The effective $|{\bf k}|^2$ plot for the $\Xi^-\Xi^-$
    channel with $t_J=3$ and the fit to the plateau.  }
  \label{fig:XiXi-k2-BOT}
\end{figure}

The only $s=-4$ systems of two octet-baryons are the $I=0,1$ $\Xi\Xi$
combinations and here we focus on $I=1$ ($\Xi^-\Xi^-$). The low-lying
states in the lattice volume that couple to the $\Xi^-\Xi^-$
interpolating operator are expected to be describable in terms of a
single-channel elastic-scattering amplitude as there are no other
states composed of two octet baryons that can couple to them.  The
GEMPs for this channel are shown in fig.~\ref{fig:XiXi-BOT},
and the resulting effective $|{\bf k}|^2$ plot is shown in
fig.~\ref{fig:XiXi-k2-BOT}.
The results obtained by fitting to the effective $|{\bf k}|^2$ plot
are shown in Table~\ref{tab:LamLamresults}, and presented graphically
in fig.~\ref{fig:LamLam-INVpcotVk2}, normalized to the pion mass.  The
effective $|{\bf k}|^2$ plot shows the downward fluctuation seen in
other correlators at time-slice $t=29$, and the resulting fit is
consistent with zero at the $2\sigma$-level.  We conclude that the
$\Xi^-\Xi^-$ interactions at this value of the pion mass are quite
weak.

\section{Statistical Scaling and Noise in Correlation Functions}
\label{sec:stat-behav}

\noindent The precision of lattice QCD calculations of any quantity is
limited by the statistical noise in the relevant correlation
functions.  Until recently~\cite{Beane:2009ky,Beane:2009gs}, the lore
has been (based on general arguments by Lepage \cite{Lepage:1989hd})
that the signal-to-noise ratios in (multi-)baryon correlation
functions degrade exponentially with time and with the number of
baryons in the system.  In Refs.~\cite{Beane:2009ky,Beane:2009gs}, we
showed that, while this behavior is observed at large times, at
intermediate times the signal-to-noise ratio does not degrade
exponentially, and in fact it is found to be independent of time for a
significant number of time slices (the ``Golden Window'') for the
sources that are used.  Further, in this window of time-slices, the
signal-to-noise ratio is essentially independent of the number of
baryons in the system.

After reviewing Lepage's arguments regarding the general behavior of
signal-to-noise ratio, and their generalization to the temporal
boundary conditions that are used in the present work, a thorough
exploration of the noise in the various correlators of the two-baryon
sector is presented.

\subsection{Aspects of the noise correlation functions and the
  signal-to-noise ratio}
\label{sec:noise-correlations}
\noindent
On gauge-field configurations with infinite temporal extent,
correlation functions of one or more baryons exhibit statistical noise
at large-times that increases exponentially with Euclidean
time~\cite{Lepage:1989hd}.  In the case of a source that has the
quantum numbers of a single positive parity nucleon, the correlation
function has the form
\begin{eqnarray}
  \langle \theta_{N}(t)\rangle &=&
  \sum_{\bf x}\ 
  \Gamma_+^{\beta\alpha}
  \ \langle 0 | \ N^\alpha({\bf x},t) \overline{N}^{\beta} ({\bf 0},0)
  \ |0\rangle
  \ \rightarrow\ Z_N \ e^{-M_N t}
  \ \ ,
  \label{eq:Gfunproton}
\end{eqnarray}
where $N^\alpha({\bf x},t)$ is an interpolating field (composed of
three quark operators) that has non-vanishing overlap with the
nucleon, $\Gamma_+$ is a positive energy projector, and the angle
brackets indicate statistical averaging over measurements on an
ensemble of configurations.  The variance of this correlation function
is given by
\begin{eqnarray}
  {\rm N}\  \sigma^2 & \sim & 
  \langle \theta^{\dagger}_N(t)
  \theta_N(t)\rangle  - \langle \theta_{N}(t) \rangle^2 \nonumber \\
  & = & 
  \sum_{\bf x,y} \Gamma_+^{\delta\alpha}\Gamma_+^{\gamma\beta\dagger}\ 
  \langle 0|\ N^\alpha({\bf x},t) \overline{N}^{\beta}({\bf y},t) N^\gamma({\bf 0},0)
  \overline{N}^{\delta}({\bf 0},0) 
  \ |0 \rangle\ \ -\ \langle \theta_{N}(t) \rangle^2
  \nonumber\\
  & \rightarrow & 
  Z_{N\overline{N}} e^{-2 M_N t} - Z_N^2  e^{-2M_N t}
  \ +\ 
  Z_{3\pi}\  e^{-3 \mpi t}\ +\ ...
  \ \ \stackrel{t\to\infty}{\rightarrow} \ Z_{3\pi}\  e^{-3 \mpi t}
  \ \ ,
  \label{eq:GGdaggerfunproton}
\end{eqnarray}
where all interaction energies have been neglected, and N is the
number of (independent) measurements (distinct from the nucleon field
operator $N$).  At large times, the noise-to-signal ratio consequently
behaves as~\cite{Lepage:1989hd}
\begin{eqnarray} {\sigma\over\overline{x}} & = & {\sigma (t)\over
    \langle \theta(t) \rangle } \sim {1\over \sqrt{\rm N}} \ e^{\left(
      M_N - {3\over 2} \mpi\right) t} \ \ .
  \label{eq:NtoSproton}
\end{eqnarray}
More generally, for a system of $A$ nucleons, the noise-to-signal
ratio behaves as
\begin{eqnarray} {\sigma\over\overline{x}} & & \sim {1\over \sqrt{\rm
      N}} \ e^{A \left( M_N - {3\over 2} \mpi\right) t} \ \
  \label{eq:NtoSnucleus}
\end{eqnarray}
at large times.

As we discussed in Ref.~\cite{Beane:2009gs}, the various
``Z-factors'', such as $Z_{3\pi}$, depend upon the details of the
sources and sinks interpolators that are used.  For the present
calculations, the projection onto zero-momentum final state nucleons,
introduces a $1/\sqrt{\rm Volume}$ suppression of the amplitudes of
the various components (except for $N\overline{N}$) in addition to
color and spin rearrangement suppressions that exists independent of
the spatial structure of the source. As a consequence, an interval of
time slices exists at short times (the ``Golden Window'') in which the
variance of the correlation function is dominated by the terms in
eq.~(\ref{eq:GGdaggerfunproton}) that behave as $\sim e^{-2 M_N t} $.
In this window, the signal-to-noise ratio of the single baryon
correlation function is independent of time.  Further, the
signal-to-noise ratio does not degrade exponentially faster in
multi-baryon correlation functions than in single-baryon correlation
functions in the ``Golden Window'' \cite{Beane:2009gs}.

The finite temporal extent introduces backward propagating states
(thermal states) into the correlation functions which lead to
exponentially worse signal-to-noise ratios at large
times~\cite{Beane:2009ky,Beane:2009gs}. These contributions are
suppressed by at least $\exp(m_\pi T)$, however, in the present work
(where $m_\pi T\sim9$), these effects cause complications. We note
that the impact of these states can be mitigated by working at larger
temporal extents and exponentially large computational resources are
not required to remove this effect.

With the high statistics that have been accumulated in the present
work, the behavior of the signal-to-noise ratio can be carefully
examined.  It is useful to form the effective noise-to-signal
plot~\cite{Beane:2009ky}, in analogy with the GEMPs.  On each time
slice, the quantity
\begin{eqnarray} {\cal S}(t) & = & {\sigma (t)\over\overline{x}(t)} \
  \ \ ,
  \label{eq:stondefn}
\end{eqnarray}
is formed, from which the energy governing the exponential behavior
(the signal-to-noise energy-scale) can be extracted via
\begin{eqnarray}
  E_s(t;t_J) & = & 
  {1\over t_J}\ \log\left({ {\cal S}(t+t_J) \over {\cal S}(t)}\right)
  \ \ \ .
  \label{eq:Estondefn}
\end{eqnarray}
For a correlation function that is dominated by a single state with a
corresponding variance correlation function dominated by a single
energy scale, the quantity $E_s(t;t_J)$ will be independent of both
$t$ and $t_J$.

\subsection{Measured signal-to-noise ratios in the one and two nucleon
  sectors}
\label{sec:S2N-N-measured}
\noindent
In the single nucleon sector, we expect that the energy scales $E_s
\sim 0$, $M_N-{3\over 2} m_\pi$, and others, contribute to the
signal-to-noise ratio.  At times when the nucleon correlation function
is in the ground state, and the variance correlation function is
dominated by the nucleon-antinucleon state, $E_s = 0$ should dominate
the signal-to-noise ratio. At large times the variance correlation
function is dominated by the 3-pion state and $E_s=M_N-{3\over 2}
m_\pi$ should dominate.  This is modified by the finite temporal
direction \cite{Beane:2009ky} as the hadrons produced by the sources
of the correlation function and the variance correlation function can
propagate forward and backward in time.
\begin{figure}[!th]
  \centering
  \includegraphics[width=1.0\columnwidth]{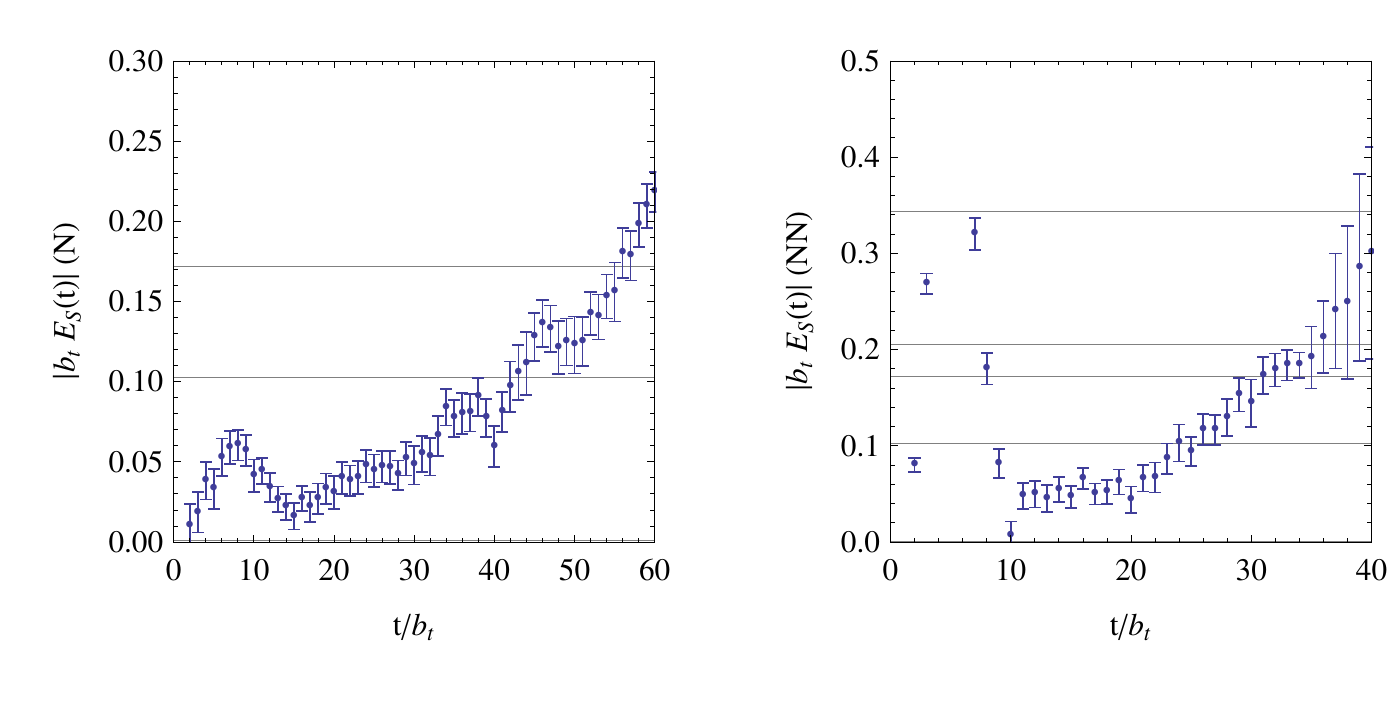}
  \caption{The energy scale of the signal-to-noise ratio in the
    nucleon (left panel) and proton-proton (right panel) correlation
    functions, as defined in eq.~(\protect\ref{eq:Estondefn}), with
    $t_J=6$.  The horizontal lines in the left panel correspond to
    $E_s=0$, $M_N-{3\over 2} m_\pi$ and $M_N-{1\over 2} m_\pi$, while
    those in the right panel correspond to $E_s=0$, $M_N-{3\over 2}
    m_\pi$, $M_N-{1\over 2} m_\pi$, $2 M_N-3 m_\pi$, $2 M_N- m_\pi$.
  }
  \label{fig:NN-noise}
\end{figure}
The measured energy scale of the signal-to-noise ratio of the single
nucleon correlation function is shown in fig.~\ref{fig:NN-noise}.  It
exhibits behavior that is consistent with expectations, and exceeds
the long-time behavior expected from the Lepage argument at
approximately time-slice $t=50$ due to the temporal boundary
conditions.

On configurations with infinite temporal extent, the proton-proton
correlation function is of the form (neglecting interactions between
the hadrons)
\begin{eqnarray}
  \langle \theta_{NN}(t)\rangle &=&
  \sum_{\bf x, y}\ 
  \Gamma_+^{\alpha\gamma\beta\rho}
  \ \langle 0 | \ N^\alpha({\bf x},t) N^\gamma({\bf y},t)
  \overline{N}^{\beta}({\bf 0},0) 
  \overline{N}^{\rho} ({\bf 0},0)
  \ |0\rangle \nonumber\\
  & \rightarrow & Z_{NN} \ e^{-2 M_N  t} + \ldots,
  \label{eq:GfunNN}
\end{eqnarray}
and the variance correlation function has the form
\begin{eqnarray} {\rm N}\ \sigma^2 &\sim& \langle
  \theta^{\dagger}_{NN}(t)
  \theta_{NN}(t)\rangle  - \langle \theta_{NN}(t) \rangle^2 \nonumber \\
  &= & \sum_{\bf x,y,z,w} \Gamma_+^{\alpha\rho\delta\psi}
  \Gamma_+^{\beta\eta\gamma\zeta\dagger}\ \langle 0|\ N^\alpha({\bf
    x},t) N^\rho({\bf y},t) \overline{N}^{\beta}({\bf z},t)
  \overline{N}^{\eta}({\bf w},t)\;\times \nonumber\\
  &&\qquad \quad\qquad\qquad\qquad \overline{N}^{\delta}({\bf 0},0)
  \overline{N}^{\psi}({\bf 0},0) N^\gamma({\bf 0},0) N^\zeta({\bf
    0},0) \ |0 \rangle\ -\ \langle \theta_{NN}(t) \rangle^2
  \nonumber\\
  & \rightarrow & Z_{NN\overline{N}\overline{N}} e^{-4 M_N t} -
  Z_{NN}^2 e^{-4 M_N t} \ +\ Z_{3\pi N\overline{N}}\ e^{-(2 M_N + 3
    \mpi) t} + Z_{6\pi}\ e^{-6 \mpi t}+\ldots
  \nonumber\\
  & \rightarrow & Z_{6\pi}\ e^{-6 \mpi t}.
  \label{eq:GGdaggerfunNN}
\end{eqnarray}
Therefore, we anticipate finding energy scales of approximately
$E_s=0, M_N-{3\over 2} m_\pi$ and $2 M_N- 3m_\pi$ in the
signal-to-noise ratio on gauge-field configurations of infinite
temporal extent.  The temporal boundary conditions imposed in the
present calculation introduce additional energy scales due to the
backward propagating states.

\begin{figure}[!th]
  \centering
  \includegraphics[width=1.0\columnwidth]
  {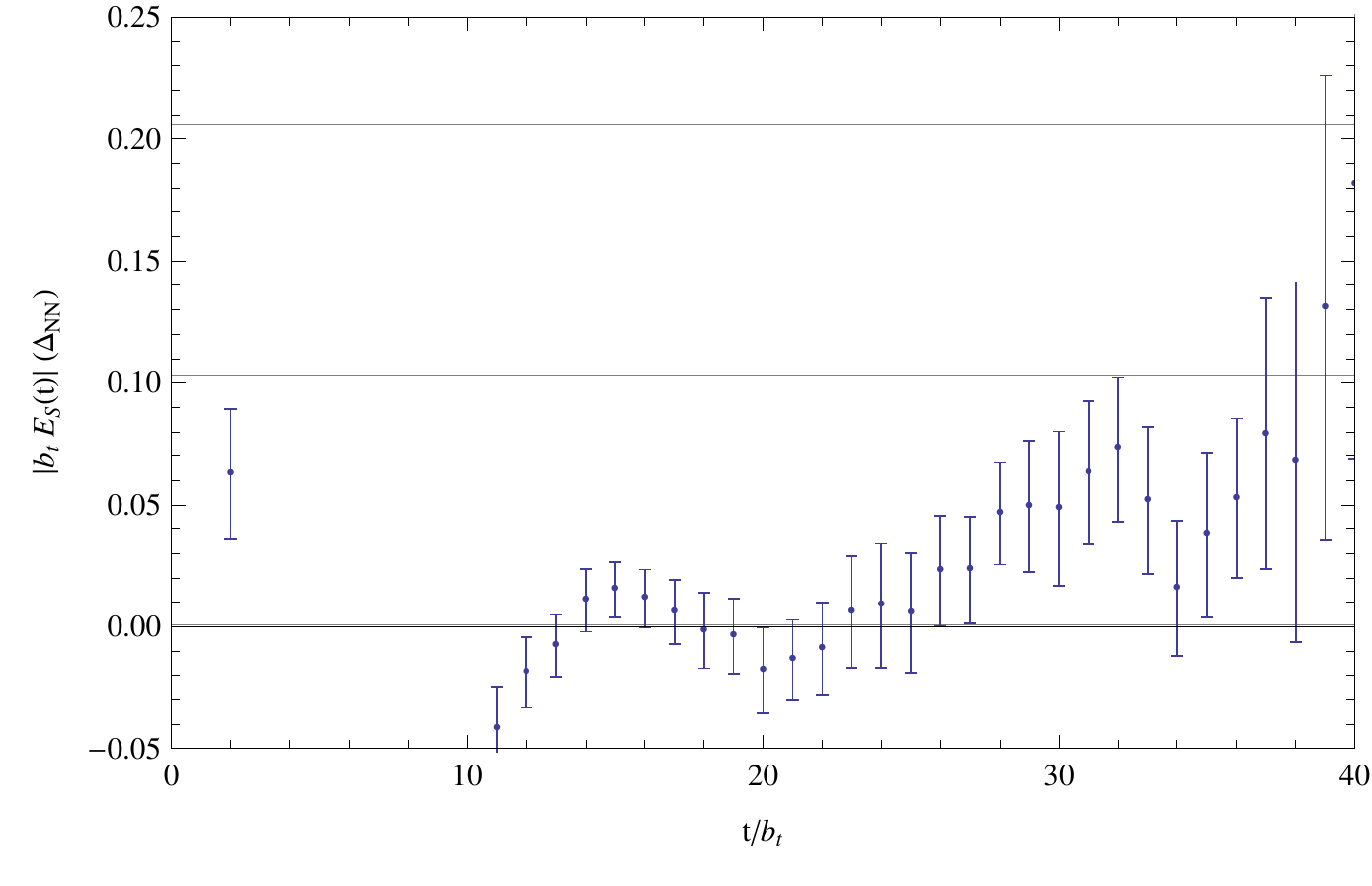}\\
  \caption{The energy scale of the signal-to-noise ratio, as defined
    in eq.~(\protect\ref{eq:Estondefn}), in the ratio of correlation
    functions that produces the shift in energy between two
    interacting protons and two isolated protons, with $t_J=6$.  The
    horizontal lines correspond to $E_s=0$, $M_N-{3\over 2} m_\pi$ and
    $2 M_N-3 m_\pi$.  }
  \label{fig:DIFF-NN-noise}
\end{figure}
Figure~\ref{fig:DIFF-NN-noise} shows the energy scale associated with
the signal-to-noise ratio for the ratio of correlation functions that
provides the energy splitting between two interacting protons and two
isolated protons from which the $p\cot\delta(p)$ is extracted.  It is
clear that the energy scale of the energy splitting is significantly
less than for the individual energies, and is consistent with zero
throughout much of the Golden Window of time slices.  This indicates
that the signal-to-noise ratio associated with the energy splitting in
the proton-proton sector, and hence the scattering parameters and
bound-state energies, are time independent for the sources and sinks
used here, and therefore do not degrade exponentially with time.  This
is an exceptionally important result, as it means that the extraction
of NN, and more generally, multi-nucleon interactions, does not
require an exponentially large number of measurements for each
relevant correlation function.  Further exploration and discussion of
this point can be found in our work on three baryon
systems~\cite{Beane:2009gs}.

\subsection{Measured signal-to-noise ratios in the $N\Sigma$ sector}
\label{sec:S2N-NSigma-measured}
\begin{figure}[!th]
  \centering
  \includegraphics[width=1.0\columnwidth]{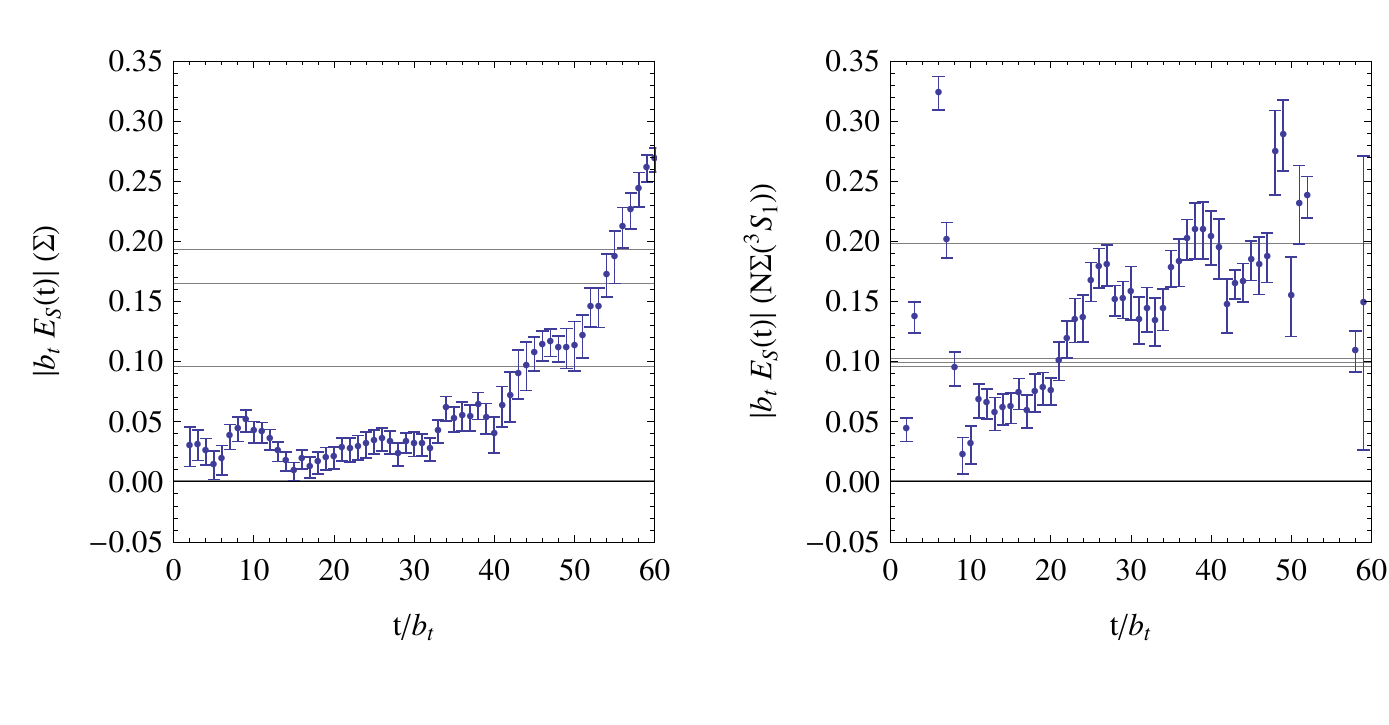}
  \caption{The energy scale of the signal-to-noise ratio in the
    $\Sigma$ (left panel) and the $n\Sigma^-$ $(^3S_1)$ (right panel)
    correlation functions, as defined in
    eq.~(\protect\ref{eq:Estondefn}), with $t_J=6$.  The horizontal
    lines in the left panel correspond to $E_s=0$,
    $M_\Sigma-m_K-{1\over 2} m_\pi$, $M_\Sigma-m_K + {1\over 2} m_\pi$
    and $M_\Sigma - m_\pi$, while those in the right panel correspond
    to $E_s=0$, $ M_N-{3\over 2} m_\pi$, $M_\Sigma - m_K - {1\over 2}
    m_\pi$, ${1\over 2} M_N + {1\over 2} M_\Sigma - {1\over 2} m_K -
    m_\pi$ and $ M_N + M_\Sigma - m_K - 2 m_\pi$.  }
  \label{fig:SN-noise}
\end{figure}

\noindent
The discussion of the behavior of the signal-to-noise ratio in the YN
sector parallels that in the multi-nucleon sector in
section~\ref{sec:S2N-N-measured}.  An important difference is the
presence of the strange quark, and the associated strange hadrons.
The lowest energy scale contributing to the signal-to-noise ratio
(beyond $E_s=0$) is $E_s = M_N-{3\over 2} m_\pi$, and so it is
expected that the degradation of the signal-to-noise ratio will be
similar to that found in the NN correlations functions.  It is found
that the associated energy scale in the $n\Sigma^-$ $(^3 S_1)$
channel, as shown in fig.~\ref{fig:SN-noise}, is somewhat less than
$M_N-{3\over 2} m_\pi$ over a number of time-slices, and starts to
exceed this value for time-slices greater than $t\gsim 20$.

\begin{figure}[!th]
  \centering
  \includegraphics[width=1.0\columnwidth]{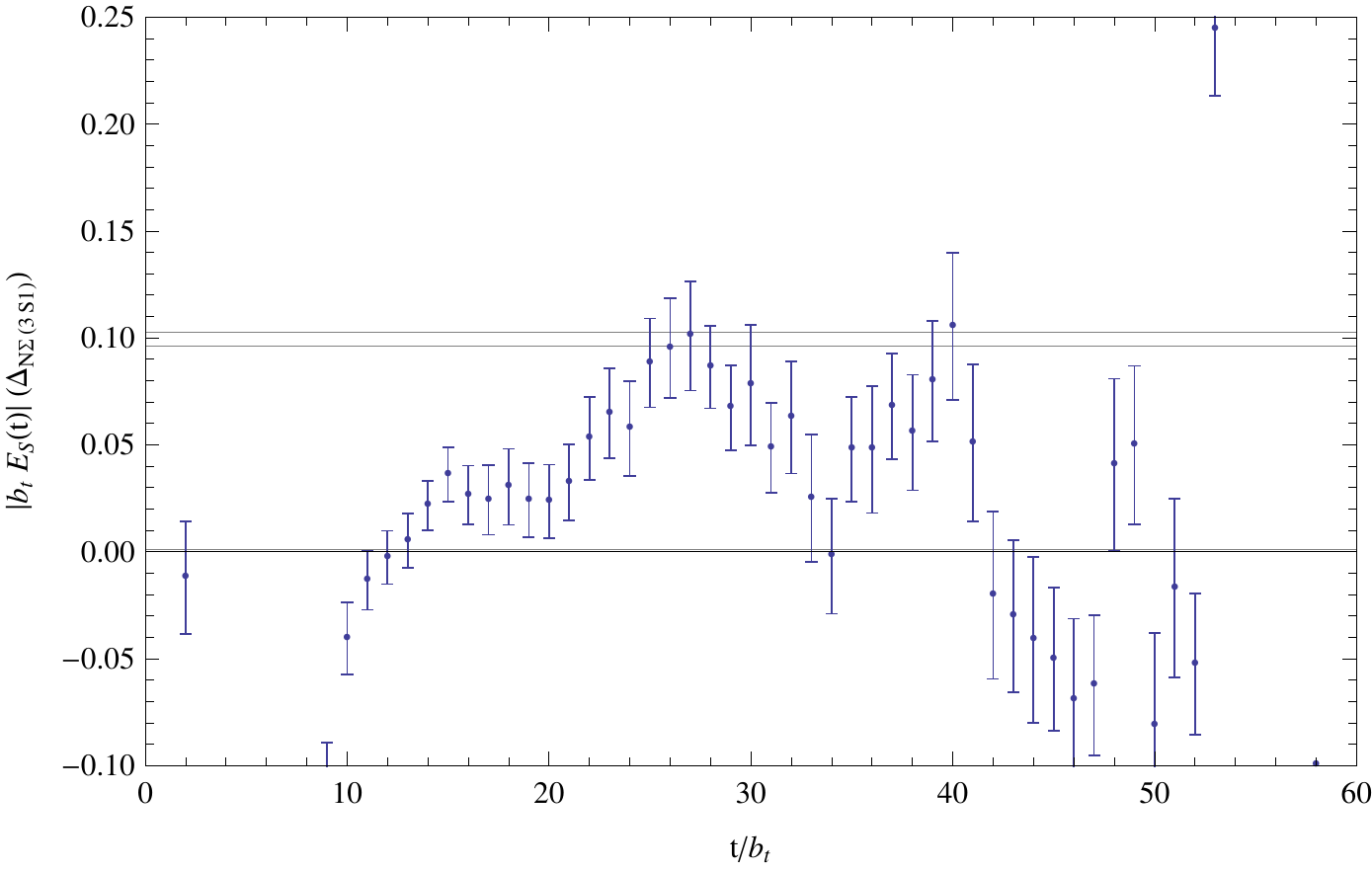}\\
  \caption{The energy scale of the signal-to-noise ratio, as defined
    in eq.~(\protect\ref{eq:Estondefn}), in the ratio of correlation
    functions that produces the shift in energy between interacting
    $\Sigma$'s and neutrons in the $^3S_1$-channel and isolated
    $\Sigma$'s and neutrons, with $t_J=6$.  The horizontal lines
    correspond to $E_s=0$, $M_N-{3\over 2} m_\pi$ and $M_\Sigma-m_K -
    {1\over 2} m_\pi$.  }
  \label{fig:DIFF-NSig3S1-noise}
\end{figure}

The energy scale associated with the signal-to-noise ratio in the
difference in energy between $N\Sigma$ interacting in the
$^3S_1$-channel and $M_N+M_\Sigma$ is shown in
fig.~\ref{fig:DIFF-NSig3S1-noise}.
The energy scale seems to be somewhat larger than in the NN sector,
and is non zero throughout the Golden Window.  The interaction is
strong in this channel, and therefore a non-zero value of $E_s$ in the
plateau region is not surprising.

\subsection{Measured signal-to-noise ratios in the one and two $\Xi$
  sectors}
\label{sec:S2N-Xi-measured}
\begin{figure}[!th]
  \centering
  \includegraphics[width=1.0\columnwidth]{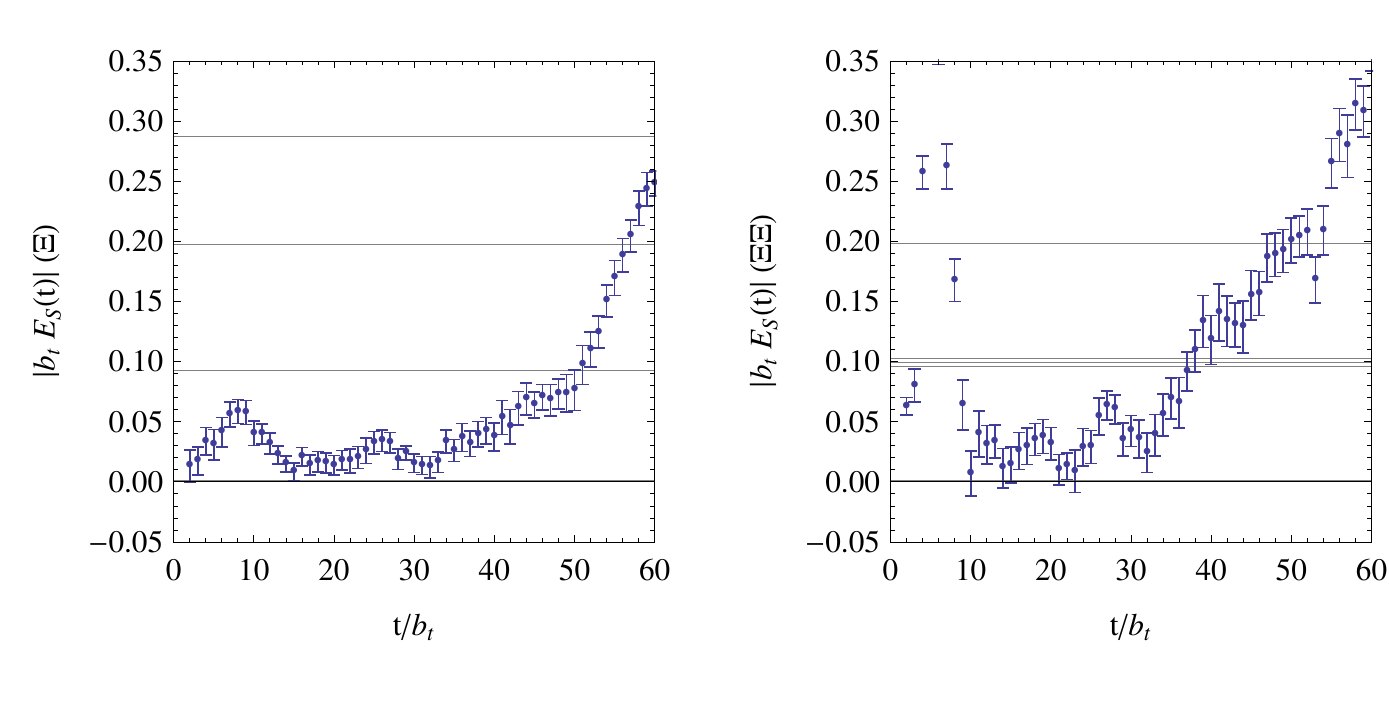}
  \caption{The energy scale of the signal-to-noise ratio in the $\Xi$
    (left panel) and $\Xi^-\Xi^-$ (right panel) correlation functions,
    as defined in eq.~(\protect\ref{eq:Estondefn}), with $t_J=6$.  The
    horizontal lines in the left panel correspond to $E_s=0$,
    $M_\Xi-m_K-{1\over 2} m_\eta$, $M_\Xi-m_K+{1\over 2} m_\eta$ and
    $M_\Xi+m_K-{1\over 2} m_\eta$, while those in the right panel
    correspond to $E_s=0$, $M_\Xi-m_K-{1\over 2} m_\eta$,
    $M_\Xi-m_K+{1\over 2} m_\eta$, $M_\Xi+m_K-{1\over 2} m_\eta$ and
    $2 (M_\Xi-m_K-{1\over 2} m_\eta)$.  }
  \label{fig:XiXi-noise}
\end{figure}

\begin{figure}[!th]
  \centering
  \includegraphics[width=1.0\columnwidth]{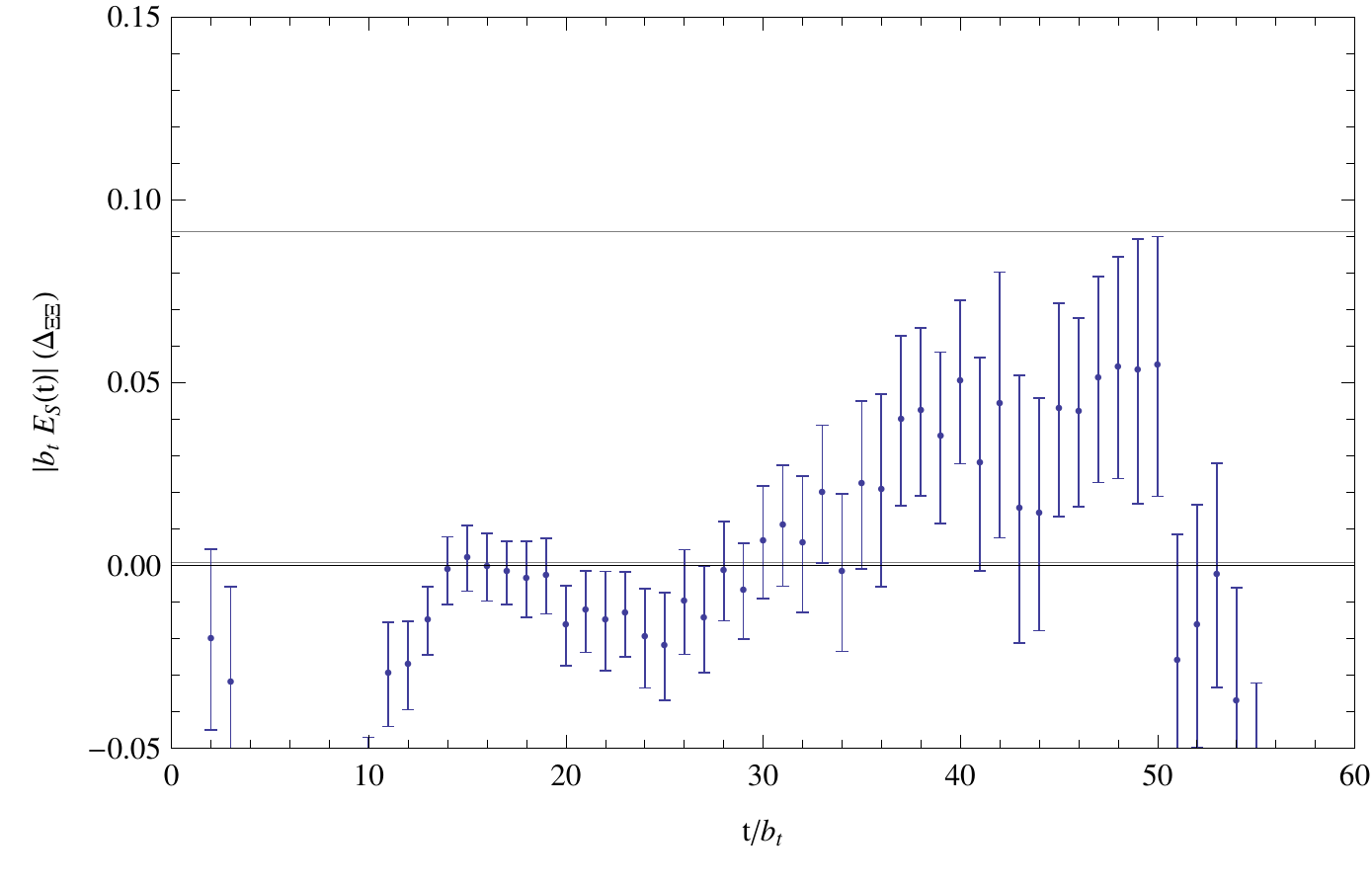}\\
  \caption{The energy scale of the signal-to-noise ratio, as defined
    in eq.~(\protect\ref{eq:Estondefn}), in the ratio of correlation
    functions that produces the shift in energy between interacting
    $\Xi^-$'s and isolated $\Xi$'s, with $t_J=6$.  The horizontal
    lines correspond to $E_s=0$ and $M_\Xi-m_K - {1\over 2} m_\eta$.
  }
  \label{fig:DIFF-XiXi-noise}
\end{figure}

\noindent
The signal-to-noise ratio in the $\Xi\Xi$ sector is noticeably better
than in the NN and the YN sectors.  The energy scale associated with
the $\Xi$ and $\Xi^-\Xi^-$ correlation functions are shown in
fig.~\ref{fig:XiXi-noise}.
The lowest energy scale contributing to the signal-to-noise ratio in
the single $\Xi$ correlation function (beyond $E_s=0$) is $E_s =
M_\Xi-m_K-{1\over 2} m_\eta$.  It is clear from the left panel in
fig.~\ref{fig:XiXi-noise} that the scale is much lower than this until
time-slice $t\sim 50$.  The situation is similar in the $\Xi^-\Xi^-$
correlation function in which the energy scale associated with the
signal-to-noise ratio is found to be much less than the anticipated
$E_s = M_\Xi-m_K-{1\over 2} m_\eta$ (beyond $E_s=0$) until time-slice
$t\sim 40$.  Further, as shown in fig.~\ref{fig:DIFF-XiXi-noise}, the
energy scale associated with the energy splitting between the
$\Xi^-\Xi^-$ state and two isolated $\Xi^-$'s is very small, and
consistent with zero, for many time-slices below $t\lsim 35$, and
increases slowly beyond this.

It is likely that the improved signal-to-noise behavior in the
$\Xi\Xi$ sector is due to a reduced overlap of the source onto the
multi-meson intermediate states in the variance correlation function
compared to purely baryonic intermediate states.  Such a reduction is
expected based on the fact that the volume occupied by multiple
$\Xi$'s is smaller than that of multiple nucleons, and serves to
extend the Golden Window beyond its range in nucleon correlation
functions.

\subsection{Scaling of correlation functions, energy levels and
  scattering phase shifts}
\label{sec:scal-energy-levels}
\begin{figure}[!th]
  \centering
  \includegraphics[width=1.0\columnwidth]{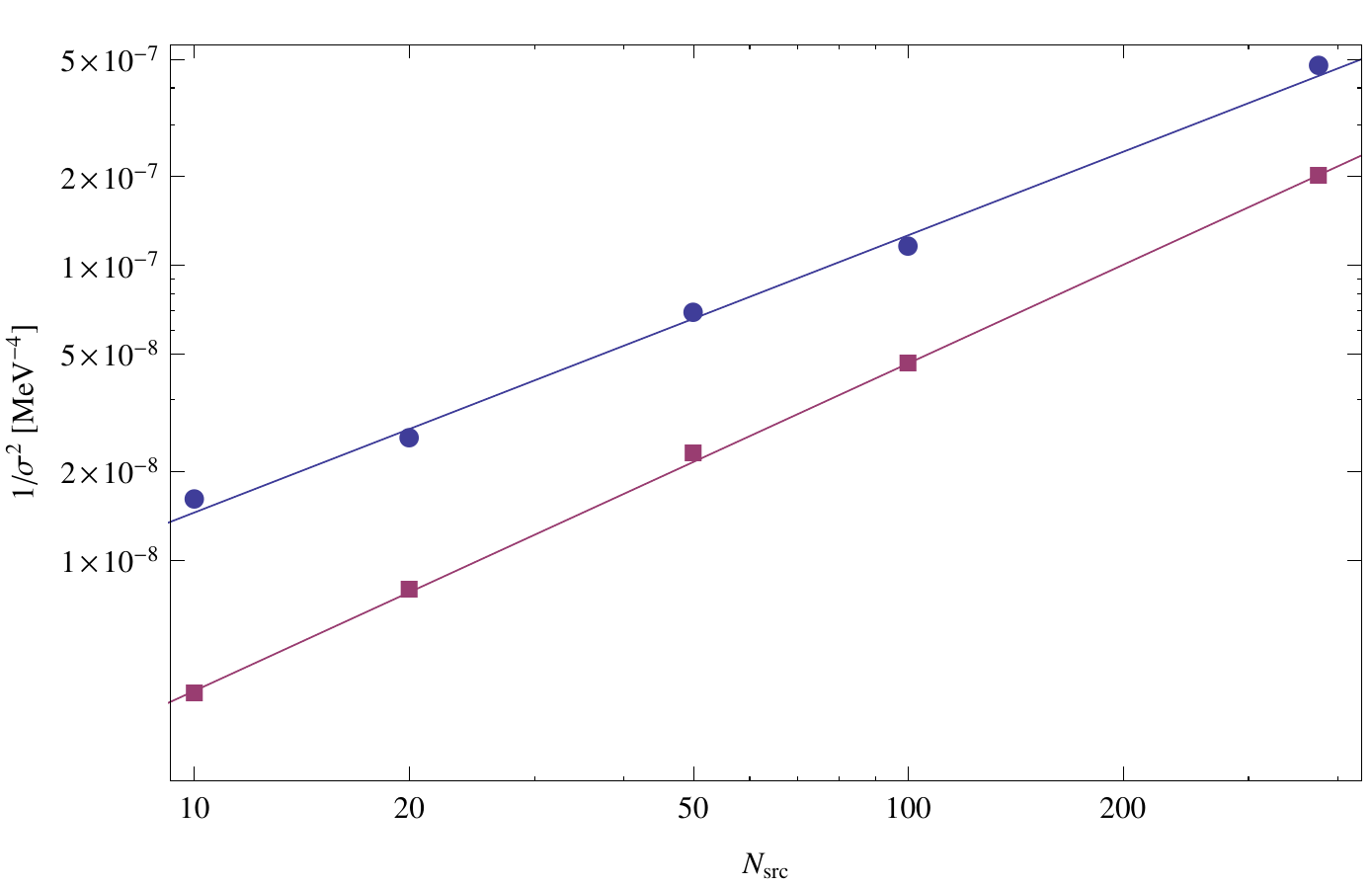}\
  \caption{The scaling of the statistical uncertainty (blue circles),
    and the statistical and fitting systematic uncertainties combined
    in quadrature (purple squares), of the extracted value of $q_0^2$,
    defined in eq.~(\protect\ref{eq:3}), for the $\Lambda\Lambda$
    system as a function of the number of measurements per
    configuration.  The vertical axis is in units of $1/{\rm MeV}^4$.
  }
  \label{fig:LamLam-Scaling-StatSys}
\end{figure}

\noindent
In generating the $\Nprops$\ measurements on this ensemble of
gauge-field configurations, we have performed an average of
\PropsperCFG\ measurements on each of the \Ncfgs\ configurations.  The
scaling of the statistical and (fitting) systematic uncertainties in
the single-hadron masses as a function of the number of measurements
and number of gauge-field configurations in this ensemble was detailed
in Ref.~\cite{Beane:2009ky}.  While the pion mass extraction was found
to saturate as the number of measurements per configuration increased,
the single baryon mass extractions did not saturate and scaled in a
way that is approximately consistent with each measurement on the
configuration being statistically independent.  It is important to
determine the scaling of uncertainties associated with scattering
parameters determined in this lattice QCD calculation because this
scaling dictates the distribution of computational resources between
the production of gauge-field configurations and the measurements
performed per configuration.  The $\Lambda\Lambda$ channel provides a
clean illustration of the scaling that is observed in the two-baryon
sector.  The fit to the blue circles in
fig.~\ref{fig:LamLam-Scaling-StatSys} shows the scaling of the
statistical uncertainty of the extracted value of $q_0^2$ from the
lowest level in the $\Lambda\Lambda$ correlation function.  The points
correspond to the inverse of the variance of $q_0^2$ as a function of
the number of sources per configuration, $N_{\rm src}$, on 1155 gauge
field configurations.  The straight line corresponds to the fit
$1/\sigma^2 = A\ N_{\rm src}^\alpha$, with $\alpha = 0.94\pm 0.04$.
The fit to the purple squares in fig.~\ref{fig:LamLam-Scaling-StatSys}
shows the scaling of the statistical and fitting-systematic
uncertainties of the extracted value of $q_0^2$ from the lowest level
in the $\Lambda\Lambda$ correlation function combined in quadrature.
The straight line fit gives $\alpha = 1.11\pm 0.02$.  It is clear that
both the statistical, and the combined statistical and systematic
uncertainties, are scaling as $\sim1/\sqrt{N_{\rm src}}$, consistent
with statistically independent measurements even up to approximately
$400$ measurements per configuration.

\begin{figure}[!th]
  \centering
  \includegraphics[width=1.0\columnwidth]{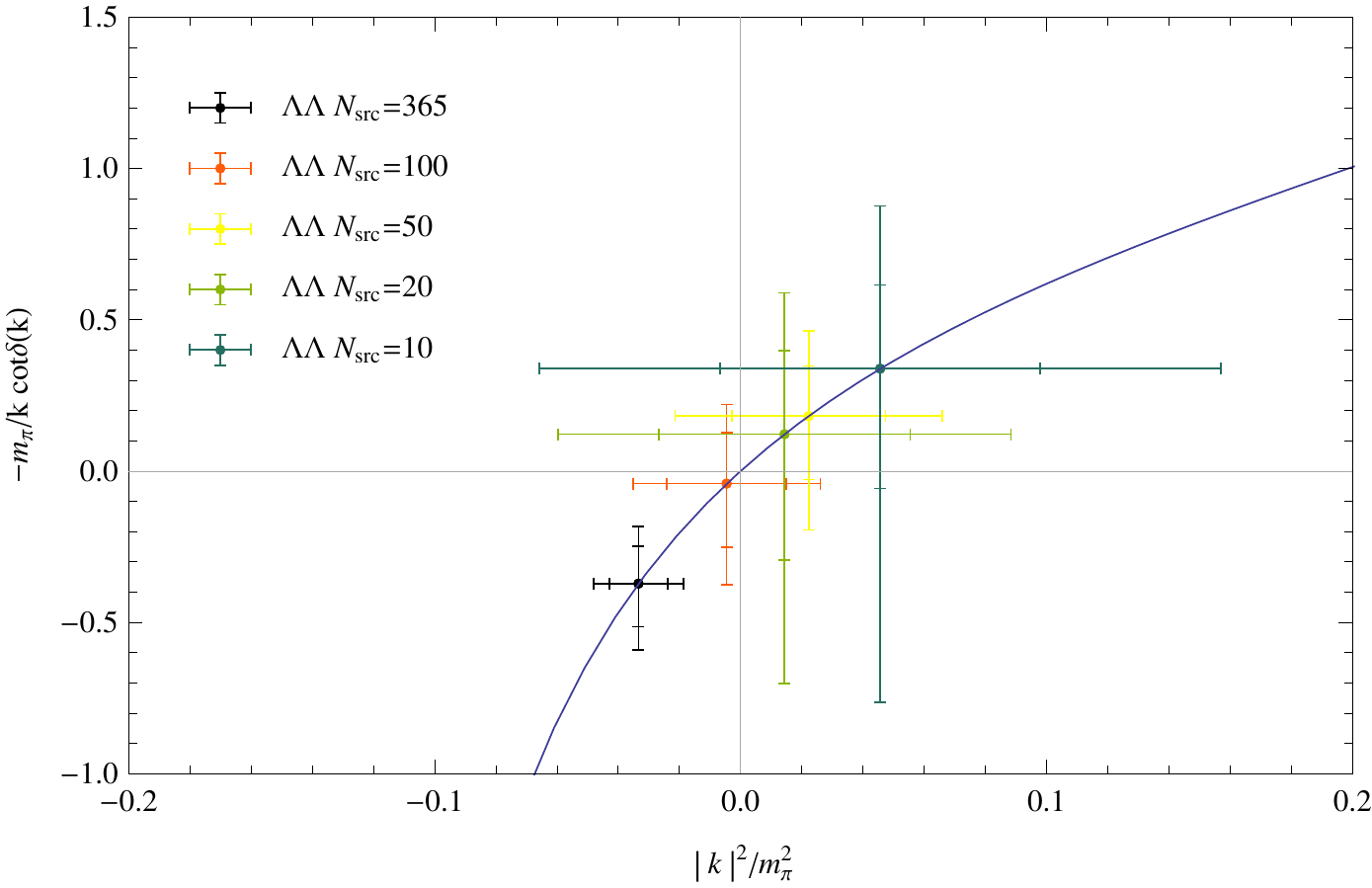}
  \caption{The extracted values of the inverse of the real part of the
    inverse of the $\Lambda\Lambda$ scattering amplitude versus the
    extracted value of $|{\bf k}|^2$, defined in
    eq.~(\protect\ref{eq:3}). The measured values correspond to
    different numbers of measurements per configuration.  }
  \label{fig:LamLam-Scaling-kcot}
\end{figure}

To emphasize the impact of the uncertainties on the extracted
scattering parameters, and in particular, the need for high-statistics
measurements of the scattering processes,
fig.~\ref{fig:LamLam-Scaling-kcot} shows the extracted values of
$(p\cot\delta)^{-1}$ for $\Lambda\Lambda$ scattering versus the
extracted value of $|{\bf k}|^2$ for different numbers of measurements
per configuration, each with $1155$ gauge-field configurations.  With
just 10 measurements per configuration the uncertainty in
$p\cot\delta$ is large enough so that it is not possible to determine
if the interaction is attractive or repulsive.  This remains the case
even for $100$ measurements per configuration.  It is only when the
number of measurements per configuration approaches $\sim~400$ that
the interaction can be determined to be attractive, but only with
$\sim 2\sigma$ significance.

Within uncertainties, the same scaling behavior is observed in all of
the two-baryon channels.  It appears that further measurements could
be performed on this ensemble of gauge-field configurations that would
continue to reduce the uncertainties in the two-baryon correlation
functions.  This statement is also valid for the single-baryon
correlation functions which do not show signs of saturation.

\section{Discussion}
\label{sec:discussion}

\begin{figure}[!th]
  \centering
  \includegraphics[width=1.0\columnwidth]{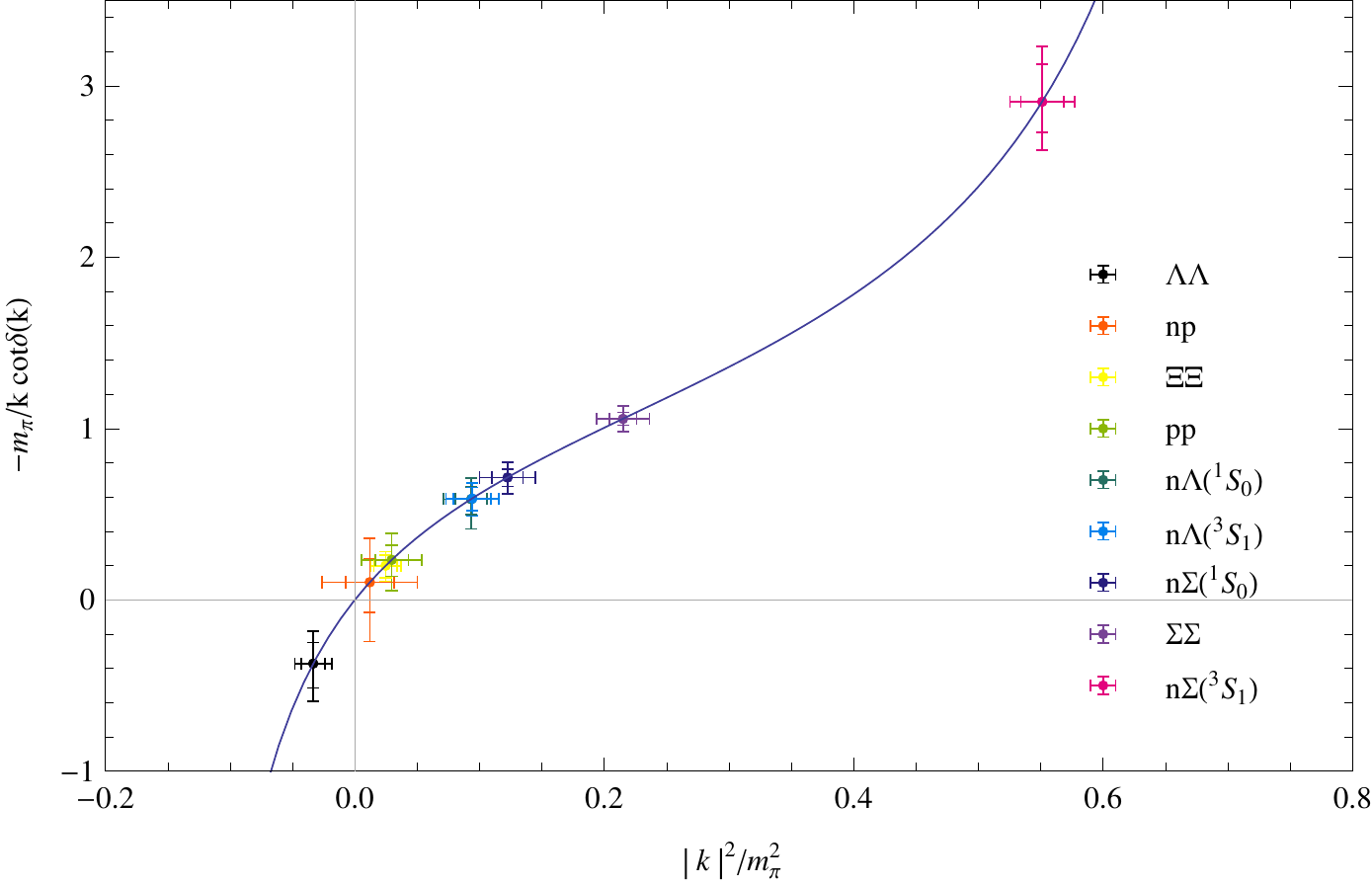}\\
  \caption{A summary of the baryon-baryon scattering information
    measured in this work.  The top-most point of the plot-legend
    corresponds to the left-most point on the plot, and the
    bottom-most point of the plot-legend corresponds to the right-most
    point on the plot.  The other points are ordered accordingly.  }
  \label{fig:All-Data}
\end{figure}

\begin{figure}[!th]
  \centering
  \includegraphics[width=1.0\columnwidth]{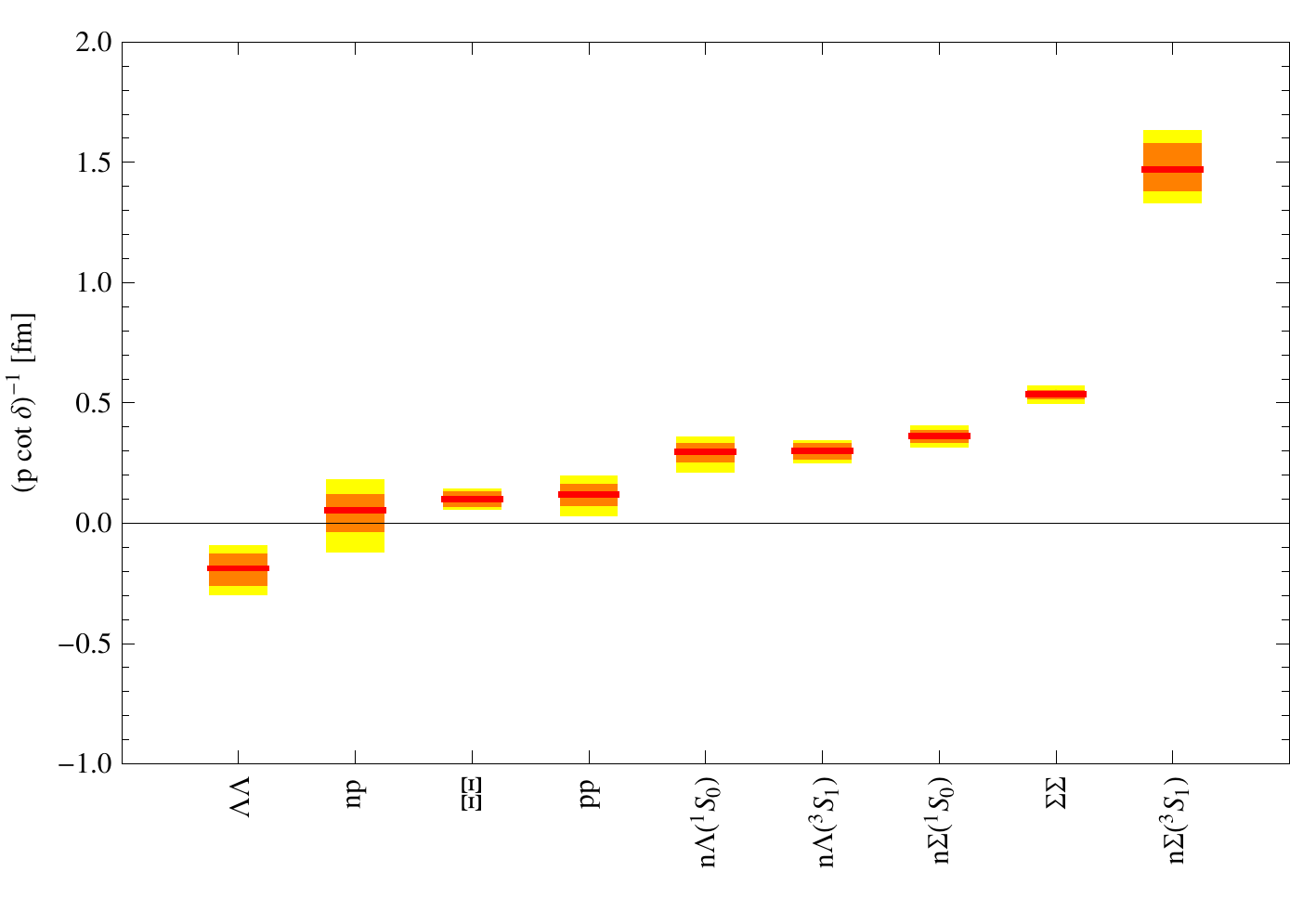}\\
  \caption{The inverse of the real part of the inverse scattering
    amplitude for all of the baryon-baryon scattering channels
    calculated in this work.  Each is determined at a different value
    of the two-baryon center-of-mass energy.  }
  \label{fig:All-Data2}
\end{figure}

\noindent
We have calculated nucleon-nucleon, hyperon-nucleon and
hyperon-hyperon interactions, with a high-statistics lattice QCD
calculation on anisotropic improved-clover gauge-field configurations
at a pion mass of $m_\pi\sim 390~{\rm MeV}$.  A summary of the
scattering information that has been extracted from the measurements
is presented in fig.~\ref{fig:All-Data}
and in fig.~\ref{fig:All-Data2}.

The phase shifts that we have obtained in the NN sector are small, and
essentially consistent with zero.  As a result we have been able to
set a tight limit on the scattering lengths in both the $^3S_1$ and
$^1S_0$ channels (without extrapolation in $|{\bf k}|^2$), as shown in
fig.~\ref{fig:NN-ALL-LQCD}.  These limits are consistent with our
previous calculations of the scattering lengths in these
channels~\cite{Beane:2006mx}, but significantly more precise.

Precise measurements in the YN sector (strangeness $= -1$) have been
obtained.  The interaction in the $n\Sigma^-$ $(^3S_1)$ channel is
found to be strong, but we are unable to determine if the interaction
is attractive or repulsive until further measurements in different
lattice volumes are performed.  Such calculations are in progress.
The measured momentum is far outside of the region for which an
effective range expansion is convergent ($|{\bf k}| < m_\pi/2$), and
consequently arguments concerning the naturalness or unnaturalness of
the scattering amplitude are not possible.  In contrast, the measured
momentum in the $n\Sigma^-$ $(^1S_0)$ channel is within the region for
which an effective-range expansion is convergent and the scattering
length and effective range are either of natural size, or there are
strong cancellations in the effective range expansion. It is clear
that the $n\Sigma^-$ interactions are strongly spin dependent, as is
expected from the long-distance contribution from one-pion
exchange. On the other hand, the $^1S_0$ and $^3S_1$ $n\Lambda$ energy
shifts are very similar, and as such we conclude that the interactions
are essentially spin independent, as would be expected from channels
without one-pion exchange.

The measurement of a negatively-shifted energy level in the
$\Lambda\Lambda$ channel (strangeness $= -2$) indicates (at the
statistical precision of the measurement) that the $\Lambda\Lambda$
interaction is attractive.  This is the only baryon-baryon channel for
which we have measured a negative energy shift.  This is an exciting
measurement as it confirms that the channel in which the H-dibaryon
\cite{Jaffe:1976yi} would arise is attractive.  The present
measurement suggest that the state does not correspond to a bound
state in the infinite-volume limit at this pion mass ($m_\pi\sim
390~{\rm MeV}$), but one can readily imagine that a bound state could
arise at a lighter pion mass.

The present work clearly demonstrates that, with sufficient
computational resources, lattice QCD can be used to extract
baryon-baryon scattering amplitudes as a function of momentum, and
hence constrain the interactions between baryons.  As has been
discussed extensively in the literature
\cite{Detmold:2007wk,Beane:2008dv}, it is not possible to directly
extract the hadron-hadron potential (unless one or more of the quarks
in each hadron is infinitely heavy), but effective interactions that
reproduce the measured scattering amplitudes can be constructed and
used in the calculation of other quantities of interest, in the same
way that the modern NN potentials are constructed to reproduce the
experimentally-measured NN scattering cross-sections.

The detailed exploration of the behavior of the signal-to-noise ratio
in the baryon-baryon correlation functions, made possible by the very
large number of measurements that have been performed, has been
exceptionally illuminating.  The importance of the ``Golden Window''
of time-slices in which the signal-to-noise ratio is essentially
independent of time cannot be overstated.  This window allows for
precise determinations of the energy-splitting between interacting
baryons and isolated baryons, and in this window, the signal-to-noise
ratio does not scale with baryon number, making precise measurements
in multi-baryon systems feasible as discussed in
Ref.~\cite{Beane:2009gs}.

This calculation is the first part of a thorough analysis of
baryon-baryon scattering at this pion mass.  Calculations in lattice
volumes that are both larger and smaller than the present lattice
volume are underway and will provide measurements of the scattering
amplitude at two additional momenta. In most channels, this will allow
for a determination of the scattering parameters (scattering lengths
and effective-range parameters) at this pion mass.  However, it is
important to keep in mind that all of these measurements will be at a
single lattice spacing, $b_s\sim 0.123~{\rm fm}$.  In order to make
precise statements, even at this larger pion mass, measurements at
smaller lattice spacings will be required.

\section{Acknowledgments}
\noindent
We thank Assumpta Parre\~no for numerous discussions.
We thank R.~Edwards and B.~Joo for help with the QDP++/Chroma
programming environment~\cite{Edwards:2004sx}. KO thanks A
Stathopoulos for collaboration in developing the EigCG algorithm used
in this work \cite{Stathopoulos:2007zi}. We also thank the Hadron
Spectrum Collaboration for the use of the anisotropic gauge-field
configurations, and extending the particular ensemble used herein.  We
gratefully acknowledge the computational time provided by NERSC
(Office of Science of the U.S. Department of Energy,
No. DE-AC02-05CH11231), the Institute for Nuclear Theory, Centro
Nacional de Supercomputaci\'on (Barcelona, Spain), Lawrence Livermore
National Laboratory, and the National Science Foundation through
Teragrid resources provided by the Texas Advanced Computing Center.
Computational support at Thomas Jefferson National Accelerator
Facility and Fermi National Accelerator Laboratory was provided by the
USQCD collaboration under {\it The Secret Life of a Quark}, a
U.S. Department of Energy SciDAC project ({\tt
  http://www.scidac.gov/physics/quarks.html}).  AT acknowledges the
kind hospitality of the NCSA.  The work of MJS and H-WL was supported
in part by the U.S.~Dept.~of Energy under Grant No.~DE-FG03-97ER4014.
The work of KO and WD was supported in part by the U.S.~Dept.~of
Energy contract No.~DE-AC05-06OR23177 (JSA) and DOE grant
DE-FG02-04ER41302. KO was also supported by NSF grant CCF-0728915. KO
and AWL were supported in part by the Jeffress Memorial Trust, grant
J-813 and DOE OJI grant DE-FG02-07ER41527. WD was also supported by
DOE OJI grant DE-SC0001784.  The work of SRB was supported in part by
the National Science Foundation CAREER grant No.  PHY-0645570. The
work of AT is currently supported by NSF Grant PHY-0555234 and DOE
grant DE-FC02-06ER41443.  Part of this work was performed under the
auspices of the US DOE by the University of California, Lawrence
Livermore National Laboratory under Contract No. W-7405-Eng-48.

%
%

\end{document}